\documentclass[aps,pra,twocolumn,tightenlines]{revtex4}

\usepackage{times,amsmath,amsfonts,amssymb,latexsym}

\usepackage{graphics,bm,bbm,multirow}
\usepackage{graphicx,epsf,epsfig}

\newcommand{\Ebf}{\mbox{\bf E}}
\newcommand{\stk}[1]{{#1}x{#1}}
\newcommand{\stak}[1]{{#1}$^{\times 2}$}
\newcommand{\csel}{C_{\rm sel}}

\newcommand{\mrm}{\rm}

\newcommand{\btab}{\begin{tabular}}
\newcommand{\etab}{\end{tabular}}

\newcommand{\ket}[1]{\mbf{|}#1\mbf{\rangle}}

\newcommand{\Scal}{{\cal S}}

\newcommand{\rank}{{\bf rank}}

\newcommand{\true}{{\mrm true}}

\newcommand{\del}{\delta}

\newcommand{\Gam}{\Gamma}

\newcommand{\eg}{\emph{e.g.}}
\newcommand{\ie}{\emph{i.e.}}

\newcommand{\bquem}{\begin{quote}\begin{em}}
\newcommand{\equem}{\end{em}\end{quote}}

\newcommand{\blist}{\begin{description}}
\newcommand{\elist}{\end{description}}

\newcommand{\bquote}{\begin{quote}}
\newcommand{\equote}{\end{quote}}

\newcommand{\ben}{\begin{enumerate}}
\newcommand{\een}{\end{enumerate}}

\newcommand{\bit}{\begin{itemize}}
\newcommand{\eit}{\end{itemize}}

\newcommand{\bea}{\begin{array}}
\newcommand{\eea}{\end{array}}

\newcommand{\bds}{\begin{displaystyle}}
\newcommand{\eds}{\end{displaystyle}}

\newcommand{\Cbf}{{\mathbf C}}

\newcommand{\mbf}[1]{\mbox{\boldmath $#1$}}

\newcommand{\refeq}[1]{(\ref{eq:#1})}

\newcommand{\beqaa}{\begin{eqnarray*}}
\newcommand{\eeqaa}{\end{eqnarray*}}

\newcommand{\beqa}{\begin{eqnarray}}
\newcommand{\eeqa}{\end{eqnarray}}

\newcommand{\bc}{\begin{center}}
\newcommand{\ec}{\end{center}}

\makeatletter
\def\beq{\@ifnextchar
[{\@tempswatrue\@beq}{\@tempswafalse\@beq[]}}
\def\@beq[#1]{\begin{equation}\edef\@tmparg{#1}\ifx\@tmparg\@e
mpty \else
    \label{#1}\fi}
\newcommand{\eeq}{\end{equation}}
\makeatother
\newcommand{\trace}{{\bf Tr}}
\newcommand{\prob}[1]{{\bf Pr}\left\{#1\right\}}
\newcommand{\norm}[1]{ \left\| #1 \right\| }
\newcommand{\normone}[1]{\norm{#1}_{\ellone}}
\newcommand{\normtwo}[1]{\norm{#1}_{\elltwo}}
\newcommand{\elltwo}{{\ell_2}}
\newcommand{\ellone}{{\ell_1}}

\newcommand{\fchn}{\mathcal{F}}

\newcommand{\qft}{{\rm qft}}
\newcommand{\svd}{{\rm svd}}

\newcommand{\bfig}{\begin{figure}}
\newcommand{\efig}{\end{figure}}

\begin{document} 

\title{Efficient measurement of quantum dynamics via compressive sensing}

\author{A. Shabani,$^{1}$ R. L. Kosut,$^{2}$, M. Mohseni$^{3}$, H. Rabitz$^{1}$, M. A. Broome$^{4}$, M. P. Almeida$^{4}$, A. Fedrizzi$^{4}$, and A. G. White$^{4}$}
\affiliation{$^{1}$Department of Chemistry, Princeton University, Princeton, New Jersey 08544, USA, $^{2}$SC Solutions, Sunnyvale, California 94085, USA, $^{3}$Research Laboratory of Electronics, Massachusetts Institute of Technology, Cambridge, Massachusetts 02139, USA, $^{4}$Centre for Engineered Quantum Systems and Centre for Quantum Computer and Communication Technology, School of Mathematics and Physics, The University of Queensland, QLD 4072, Australia}

\begin{abstract}
The resources required to characterise the dynamics of engineered
quantum systems---such as quantum computers and quantum sensors---grow
exponentially with system size. 
Here we adapt techniques from compressive sensing to exponentially reduce
the experimental configurations required for quantum process
tomography. Our method is applicable to dynamical processes that are known
to be nearly-sparse in a certain basis
and it can be implemented using only single-body
preparations and measurements. We perform efficient, high-fidelity
estimation of process matrices on an experiment attempting to
implement a photonic two-qubit logic-gate. The data base is obtained
under various decoherence strengths. We find that our technique is
both accurate and noise robust, thus removing a key roadblock to the
development and scaling of quantum technologies.
\end{abstract}

\maketitle 

Understanding and controlling the world at the nanoscale---be it in
biological, chemical or physical phenomena---requires quantum
mechanics. It is therefore essential to characterize and monitor
\emph{realistic} complex quantum systems that inevitably interact with
typically uncontrollable environments. One of the most general
descriptions of open quantum system dynamics is a quantum
map---typically represented by a \emph{process matrix}
\cite{Nielsen:book}. Methods to identify the process matrix are
collectively known as \emph{quantum process tomography} (QPT)
\cite{Nielsen:book,altepeter2003aaq,DQPT,lobino2008ccq}. For a
$d$-dimensional quantum system, they require $\mathcal{O}(d^{4})$
experimental configurations: combinations of input states, on which
the process acts, and a set of output observables. For a system of $n$
of the simplest quantum objects, namely qubits ---two-level quantum
systems---$d{=}2^n$. The required physical resources hence scale
\emph{exponentially} with system size.
In principle, a single
generalized measurement is sufficient for full process tomography in a
extended Hilbert space relying on
 highly nonlocal many-body
measurements that are physically unavailable
\cite{Mohseni:08}. Recently, a number of alternative methods have been
developed for efficient and selective estimation of quantum processes
\cite{emerson2007scn,bendersky2008see,branderhorst2009sqp}. However,
full characterization of quantum dynamics of comparably small systems,
such as a recently demonstrated 8-qubit ion trap
\cite{haffner2005sme}, would still require over a billion experimental
configurations, clearly a practical impossibility. So far, process
tomography has therefore been limited by experimental, and---to a
lesser extent---by off-line computational resources, to systems of 2
and 3 qubits \cite{obrien2004qpt,Bialczak,monz2008rqt}.

Here we adapt techniques from \emph{compressive sensing} (CS) to
develop an \emph{experimentally efficient} method for QPT. It requires
only $\mathcal{O}(s \log d)$ configurations if the process matrix is
\emph{$s$-compressible} in some known basis, \ie, it is \emph{nearly
sparse} in that it can be well approximated by an $s$-sparse process
matrix. This is usually the case, because engineered quantum systems
aim to implement a \emph{unitary} process which is maximally-sparse in
its eigenbasis. In practice, as observed in QPT experiments in
liquid-state NMR \cite{childs2001rqp,boulant2003rme,weinstein2004qpt},
photonics \cite{mitchell2003dpp,obrien2004qpt,nambu2005ein}, ion traps
\cite{Riebe:06}, and superconducting circuits \cite{Bialczak}, a
near-unitary process will still be nearly-sparse in this basis, and
still compressible. The near sparsity emanates from decoherence
originating in few dominant system-environment interactions. This is
more apparent for weakly decohering systems
\cite{mohseni2009emp,kofman2009}.

We experimentally demonstrate our algorithm by estimating the 240 real parameters of the process matrix of a canonical photonic two-qubit gate, Fig.~\ref{fig:setup}, from a reduced number of configurations. For example, from just 18 and 32 configurations, we obtain fidelities of 94\% and 97\% with process matrices obtained from an overcomplete set of all 576 available configurations.

Compressive sensing provides methods for compression of information carried by a large-size signal into a significantly smaller one along with efficient convex optimization algorithms to decipher this information \cite{candes2008phl,candes2008rcs}. Originally developed to exploit compressible features of natural audio and video signals, applications of compressive sensing have recently found their way to quantum tomography: Simulations of compressive sensing for QPT \cite{Kosut:08qpt}, application to ghost-imaging \cite{katz2009cgi}, and
quantum state tomography for \emph{low-rank} density matrices
\cite{Gross:09}. The latter provides a quadratic reduction of physical
resources compared to standard state tomography, \ie, for a density
matrix of rank $r$, $\mathcal{O}(rd \log^2 d)$ vs. standard $d^2$
settings, and it also has the main advantage that rank is basis
independent.

Under reasonable assumptions, a quantum map on a $d$-dimensional space has the general representation \cite{Nielsen:book},
\beq[eq:osr] 
\mathcal{S}(\rho){=}\sum_{\alpha,\beta=1}^{d^2}
\chi_{\alpha\beta}\Gamma_\alpha\rho\Gamma_{\beta}^{\dagger}
\eeq
where $\chi$, the $d^2 \times d^2$ process matrix, is positive semidefinite, $\chi{\geq}0$, and trace preserving, $\sum_{\alpha,\beta} \chi_{\alpha\beta} \Gamma_\beta^{\dagger} \Gamma_\alpha {=} I_{d}$, with $\{\Gamma_{\alpha}\}$ an orthonormal matrix basis set, $\trace(\Gamma_\beta^{\dagger}\Gamma_\alpha){=}\delta_{\alpha\beta}$. Note that sparsity is a property of the map representation not the map itself.
Data is collected by preparing an ensemble of identical systems in one of the states $\{\rho_1,...,\rho_k\}$, inputting them to the process $\chi$, and then measuring an observable $M$ from the set $\{M_1,...,M_\ell\}$. For a pair ($\rho,M$), the outcome will be $ y_{M,\rho}{=}\trace(\mathcal{S}(\rho)M). $ If the experiment is repeated for all configurations, \ie,
$(\rho_i,M_i),i{=}1,\ldots,m{=}k\ell$, the relation between the vector of
outcomes $y{=}[y_{M_1,\rho_1},\ldots,y_{M_m,\rho_m}]^T$ and the true
process matrix, denoted by $\chi_0$, can be represented by a linear
map $y{=}\Phi \vec\chi_0$, where $\vec\chi_0$ is the vectorized form of
the process matrix $\chi_0$ and $\Phi$ is an $m{\times} d^{4}$ matrix of
coefficients of the form
$\trace(\Gamma_\alpha\rho_i\Gamma_{\beta}^{\dagger}M_i)/\sqrt{m}$.

In general, estimating a sparse process matrix with an unknown
sparsity pattern from an underdetermined set of linear equations
($m{<}d^4$) would seem highly unlikely.  Compressive sensing, however,
tells us that this can be done by solving for $\chi$ from the convex
optimization problem:
\beq[eq:dec] 
\mbox{\rm minimize} 
\normone{\vec\chi} \
\mbox{\rm subject to}
\normtwo{y-\Phi\vec\chi} \leq \varepsilon,
\eeq
\noindent and positive-semidefinite and trace-preserving conditions as
defined above. The parameter $\varepsilon$ quantifies the level of
uncertainty in the measurements, that is, we observe
$y{=}\Phi\chi_0{+}w$ with $\normtwo{w} \leq \varepsilon$.
From \cite{candes2008rcs,Candes:08}, recovery
via \refeq{dec} is ensured if (i) the matrix $\Phi$ satisfies the \emph{restricted isometry property}:
\begin{equation}
\label{eq:rip}
1-\del_s
\leq
\begin{displaystyle}
\frac{
\normtwo{ \Phi\vec{\chi}_1(s)-\Phi\vec{\chi}_2(s) }^2
}{
\normtwo{ \vec{\chi}_1(s)-\vec{\chi}_2(s) }^2
}
\end{displaystyle}
\leq
1+\del_s
\end{equation}
for all $s$-sparse $\chi_1(s),\ \chi_2(s)$ process matrices; (ii) the \emph{isometry constant} $\del_{2s}{<}\sqrt{2}-1$ and (iii) the number of configurations $m \geq C_0 s\log(d^4/s)$.
Under these conditions, the solution $\chi^\star$ of \refeq{dec} satisfies,
\beq[eq:perf]
\normtwo{\vec\chi^\star-\vec\chi_0}
\leq
\frac{C_1}{\sqrt{s}}
\normone{\vec\chi_0(s)-\vec\chi_0}+C_2\varepsilon
\eeq
where $\chi_0(s)$ is the best $s$-sparse approximation of $\chi_0$ and
$C_0,C_1,C_2$ are constants on the order of $\mathcal{O}(\del_s)$, see
Appendix \ref{app:rip}. The restricted isometry property states that two $s$-sparse
process matrices $\chi_1(s)$ and $\chi_2(s)$ can be distinguished if
their relative distance is nearly preserved after the measurements,
\ie, under transformation by $\Phi$. If the measurements are noise
free, $\varepsilon{=}0$, and the process matrix is actually
$s$-sparse, $\chi_0{=}\chi_0(s)$, then the right hand side of
\refeq{perf} is zero leading to perfect recovery,
$\chi^\star{=}\chi_0$. Otherwise the solution tends to the best
$s$-sparse approximation of the process matrix plus the additional
term due to measurement error $\varepsilon$. If for an $n$-qubit QPT
with $d{=}2^n$ the conditions of the above analysis are satisfied,
then the number of experimental configurations $m$ scales
\emph{linearly} with $sn$, specifically, $m {\geq} C_0s(4n\log
2{-}\log s){=}\mathcal{O}(sn)$. In the appendix, using the
measure concentration properties of random matrices, following the
arguments in \cite{candes2008rcs,Candes:08}, we show that if $\Phi$ is
constructed from random input states $\{\rho_i\}$, and random
observables $\{M_i\}$, then the restricted isometry in \refeq{rip}
holds with high probability. Also a test is presented to certify the
sparsity assumption.

\bfig[t]
\centering
\btab{c}
\epsfig{file=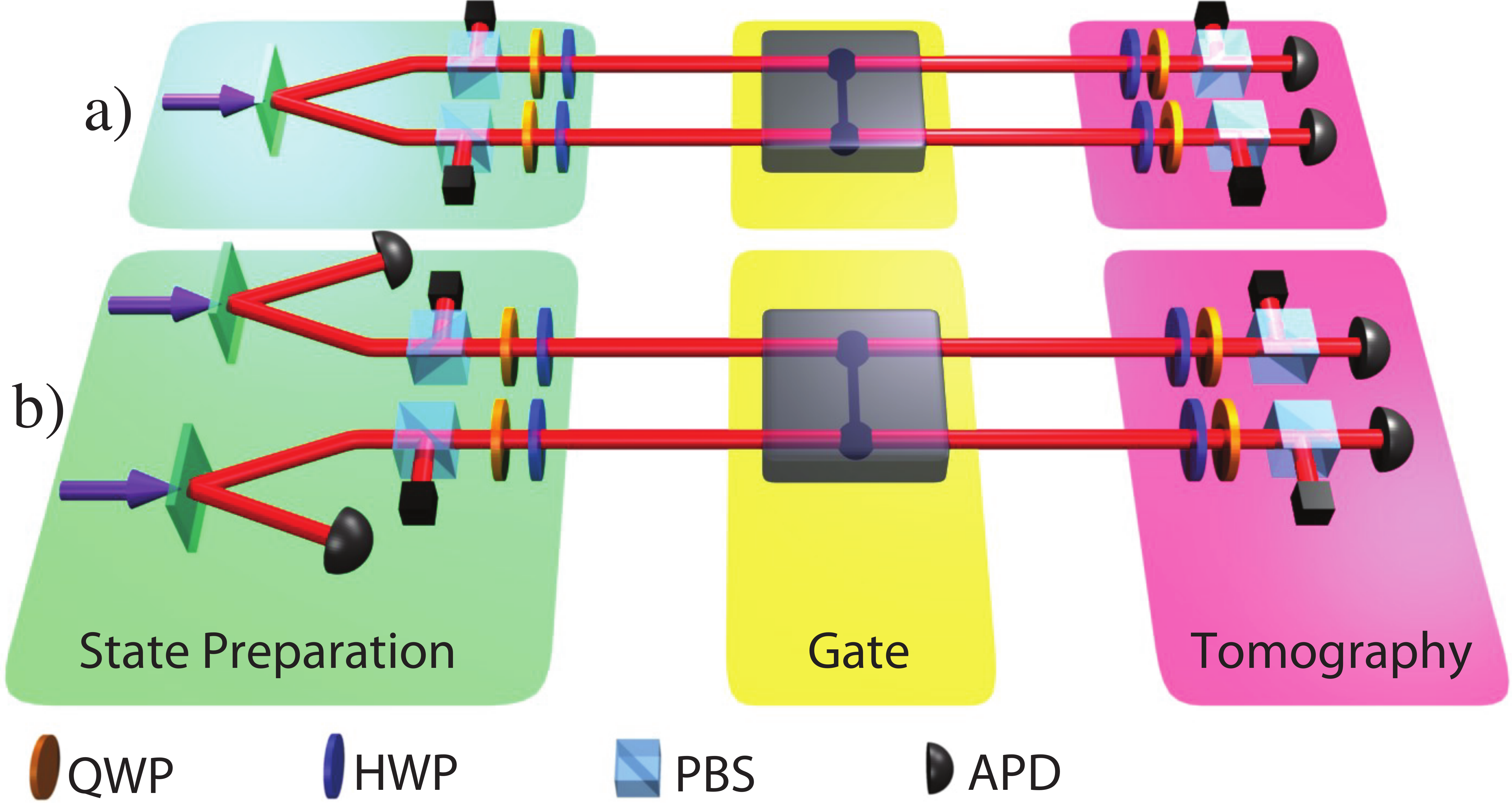,width=.9\columnwidth}
\etab
\caption{Experimental scheme. Two-photon inputs were prepared with either (a) a high-rate, non-scalable, two-photon source or (b) a low-rate, scalable, four-photon source. The qubits are encoded using polarisation, as described in the text. The quantum process is a photonic entangling-gate. A measurement configuration is defined as some combination of state preparation and tomography, implemented here with quarter- and half- waveplates (QWP, HWP), polarizers (PBS), and photon detectors (APD). For details see Appendix \ref{app:experiment}.}
\label{fig:setup}
\efig

\bfig[t]
\centering
\epsfig{file=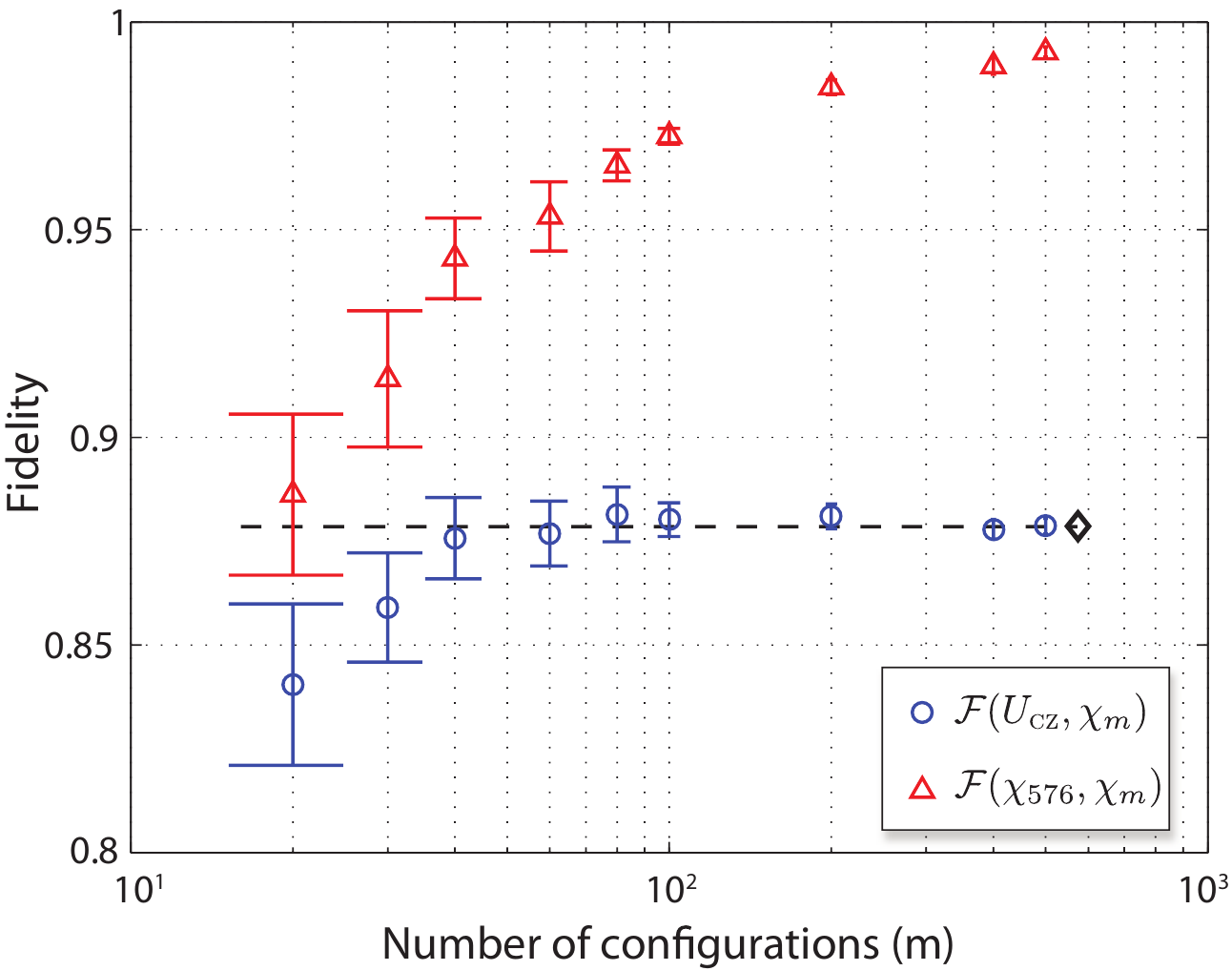,width=.9\columnwidth}
\caption{Process fidelities vs. number of input-output configurations, for each compressive QPT estimate, $\chi_{m}$, in the gate-basis of the ideal \textsc{cz}-gate for the lowest measured noise level, $\mathcal{P}{=}0.91$. The dashed line shows the fidelity of the full estimate $\mathcal{F}(U_{\textsc{cz}},\chi_{576}){=}0.89$ (black diamond). Error bars are obtained by solving \refeq{dec} for $50$ different random combinations of $m$ inputs and observables.}
\label{fig:cqpt}
\efig

\begin{figure*}[t]
  \begin{center}
\epsfig{file=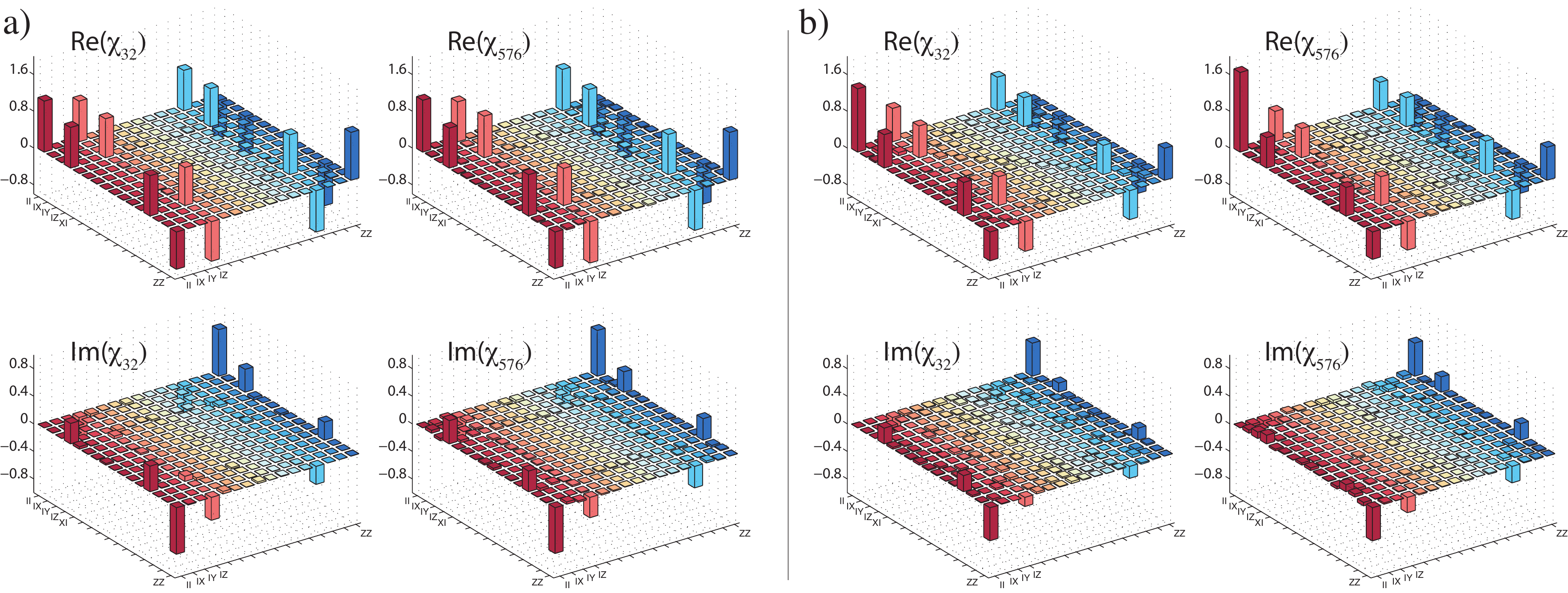,width=.9\textwidth}  
\end{center}
\vspace{-4mm}
\caption{Real and imaginary process matrix elements in the Pauli basis
for the CQPT estimate $\chi_{32}$, 32 configurations (left) vs. full data estimate $\chi_{576}$, 576 configurations (right) for A) a low noise, 2-photon experiment, $\mathcal{P}{=}0.91$, and B) a high-noise, 4-photon experiment, $\mathcal{P}{=}0.62$. The CQPT reconstructions have fidelities, $\mathcal{F}(\chi_{576},\chi_{32})$, of 95\% and 85\% respectively. The CQPT estimation accuracy is excellent for low noise, and reliable even for high noise, see Appendix \ref{app:experiment} for more details.}
\label{fig:chi32}
\end{figure*}

A nearly-sparse process matrix can thus be encoded into an \emph{exponentially smaller} number of measurement outcomes, which can be recovered to within the bounds of \refeq{perf} by solving \refeq{dec}. We now test our algorithm experimentally against standard QPT on a two-qubit gate under a range of decoherence---and thus sparsity---conditions.  We used a photonic controlled-phase, \textsc{cz}, gate, Fig.~\ref{fig:setup} where the qubits are encoded in orthogonal polarization states of single photons ($\ket{H}$, horizontal, and $\ket{V}$, vertical). We performed full process tomography \cite{mitchell2003dpp,obrien2004qpt,nambu2005ein} of the gate with both 2-photon and 4-photon arrangements, preparing 16 pair-wise combinations of the 4 input states $\{\ket{H}, \ket{V}, \ket{D}, \ket{R} \}$ and, for each input, measuring 36 two-qubit combinations of the observables $\{\ket{H},\ket{V},\ket{D}, \ket{A}, \ket{R}, \ket{L} \}$, where $\ket{D} {=} (\ket{H} {+} \ket{V})/\sqrt{2}$, $\ket{A}{=}(\ket{H}{-}\ket{V})/\sqrt{2}$, $\ket{R} {=} (\ket{H} {+} i\ket{V}/\sqrt{2}$, and $\ket{L}{=}(\ket{H}{-}i\ket{V}/\sqrt{2}$. These 576 input-output configurations represent an overcomplete set which allows the best possible estimate of the quantum process, denoted $\chi_{576}$ \cite{obrien2004qpt}.

The compressed quantum process tomography (CQPT) estimate of the $16
{\times} 16$ process matrix, denoted $\chi_m$, is obtained by solving
\refeq{dec} with $y{=}\csel p$ and $\Phi{=}\csel G$ where $p$ is the
$576\times 1$ experimental probabilities corresponding to each of the
576 configurations, $G$ is the ${576 {\times} 256}$ matrix obtained
from all the configurations and the basis set [$\Gam_\alpha$ in
\refeq{osr}],and $\csel$ is the ${m {\times} 576}$ matrix
corresponding to taking a selection of $m\leq 576$ of all possible
configurations. The basis set is obtained from the singular-value
decomposition of the ideal \textsc{cz}-gate: the process matrix in
this basis is maximally sparse with a \emph{single} non-zero
1,1-element which equals $4$. The measurement error bound
$\varepsilon$ in \refeq{dec} is chosen to be just slightly larger than
$\sqrt{m}\sigma$, where $\sigma$ is the minimum feasible
root-mean-square level obtained from \refeq{dec} using all
configurations, \ie, with $\csel{=}I_{576}$.  We quantify decoherence
using the \emph{process purity}, $\mathcal{P}{=}\trace(\chi_m^2/d^2)$,
which varies from 0 for a completely decohering channel, to 1 for a
unitary process: in our experiment we used six decoherence levels (see
Appendix \ref{app:experiment} for details), giving purities of $\{0.62, 0.74, 0.77, 0.79,
0.86, 0.91\}\pm0.01$.

Figure~\ref{fig:cqpt} shows, for the lowest decoherence level, the \emph{process fidelities} \cite{obrien2004qpt} versus the number of randomly-selected configurations, $m$. Each process matrix, $\{\chi_m\}$, is obtained by solving \refeq{dec}. We use the fidelity between (i) the compressive measurement and the ideal, $\mathcal{F}(U_{\textsc{cz}},\chi_m)$; and (ii) the compressed and optimal measurements, $\mathcal{F}(\chi_{576},\chi_m)$.
Note that as $m$ increases the fidelity with the ideal converges to the value of 0.89 obtained from $\chi_{576}$; likewise, the fidelity with the full estimate converges to unity. Similar plots exist for every level of decoherence, with fidelities reduced accordingly.

We have so far used random selections of probabilities from the full data set, which allows us a comprehensive test of compressive sensing theory. Experiments, however, don't yield probabilities but physical quantities, e.g. count rates. To date, algorithms for more efficient state \cite{Gross:09} or process tomography have assumed probabilities as a starting point. Since normalization is an issue to some extent in all physical architectures, it will be necessary to investigate the robustness and scalability of algorithms for real-world experiments. 

For our photonic two-qubit gate, which is lossy and intrinsically probabilistic, the probabilities were obtained by normalising counts using a full basis set of observables extracted from all measurements, $I_{576}$. Having sufficient configurations to allow for normalisation necessarily imposes limits on CQPT efficiency: for low $m$, we are restricted in how random our selections can be. (Details and some permissible configurations in Appendix \ref{app:normalization}). As an example, Fig.~\ref{fig:chi32} shows process matrices reconstructed via CQPT from just one of these configurations compared to the respective full data estimates. We used 32 combinations of the 16 inputs $\{\ket{H},\ket{V},\ket{D},\ket{R}\}$ and 2 observables $\{\ket{R}\ket{I},\ket{I}\ket{R}\}$, where $I$ is the identity. The agreement is excellent as one can see from the fidelities and the correct reproduction of imaginary elements---which are ideally zero. Another striking feature is that we obtain highly faithful reconstructions of a non-local process using only \emph{local} measurements \cite{Mohseni:08}. 

A further crucial test is whether CQPT enables us to locate errors and implement necessary corrections: a common example is identifying local rotations that move the process closer to the ideal. By optimising $\mathcal{F}(U_{\textsc{cz}}, \chi_{32})$, we calculated local corrections to $\chi_{32}$; applying them to the full estimate $\chi_{576}$, $\mathcal{F}(U_{\textsc{cz}}, \chi_{576})$ improved, on average, over all decoherence levels, by 4.1\%. This is very close to the average 4.9\% improvement obtained by calculating and applying local corrections \emph{directly} to $\chi_{576}$. Even a low-configuration CQPT estimate of a noisy process therefore enables improvements.

That high-fidelity estimates are obtained by CQPT can be understood
from the error bound \refeq{perf} which shows that the CQPT estimate
tends towards the best $s$-sparse approximation of the true process,
in this case our best estimate $\chi_{576}$. Fig.~\ref{fig:loglog}
shows the process matrix elements, sorted by relative magnitude, for
low and high noise levels, in two basis sets. The $s$-sparse
approximation levels indicated in \refeq{perf} are reached where the
matrix elements drop below the error threshold $(0.01$--$0.02)$. For
the corresponding $m$, we can therefore expect a successful,
high-fidelity, CQPT reconstruction. In the \textsc{cz}-basis, the
plots show that for low noise, $s{\in}[20,30]$, which correlates well
with the fidelities in Fig.~\ref{fig:cqpt}; for high noise
$s{\in}[40,60]$. Although the process matrix is still somewhat sparse
in the Pauli-basis (Fig.~\ref{fig:chi32}), the corresponding plots in
Fig.~\ref{fig:loglog} indicate that ${\sim}100$ configurations are
needed to obtain an estimate of comparable quality. Furthermore, the
sorted magnitude values in the \textsc{cz} basis decay exponentially,
which is sufficient to declare the process matrix $s$-compressible,
\eg, \cite{cosamp:08,baraniuk2010mbc}. Intriguingly, this exponential
decay is a signature of \emph{model-based compressive sensing} where
the scaling goes from $m{=}\mathcal{O}(s\log(d/s))$ to
$m{=}\mathcal{O}(s)$ \cite{baraniuk2010mbc}. This demands further
investigation, since it appears that QPT fits this framework,
particularly when the process matrix is expanded in the ideal basis
corresponding to the unitary design goal.

Our experimental results are supported numerically by simulations of a 2-qubit process, see Appendix \ref{app:sim2qb}, and of 3-, and 4-qubit processes, see Appendix \ref{app:simnqb}.

\bfig[t]
\centering
\epsfig{file=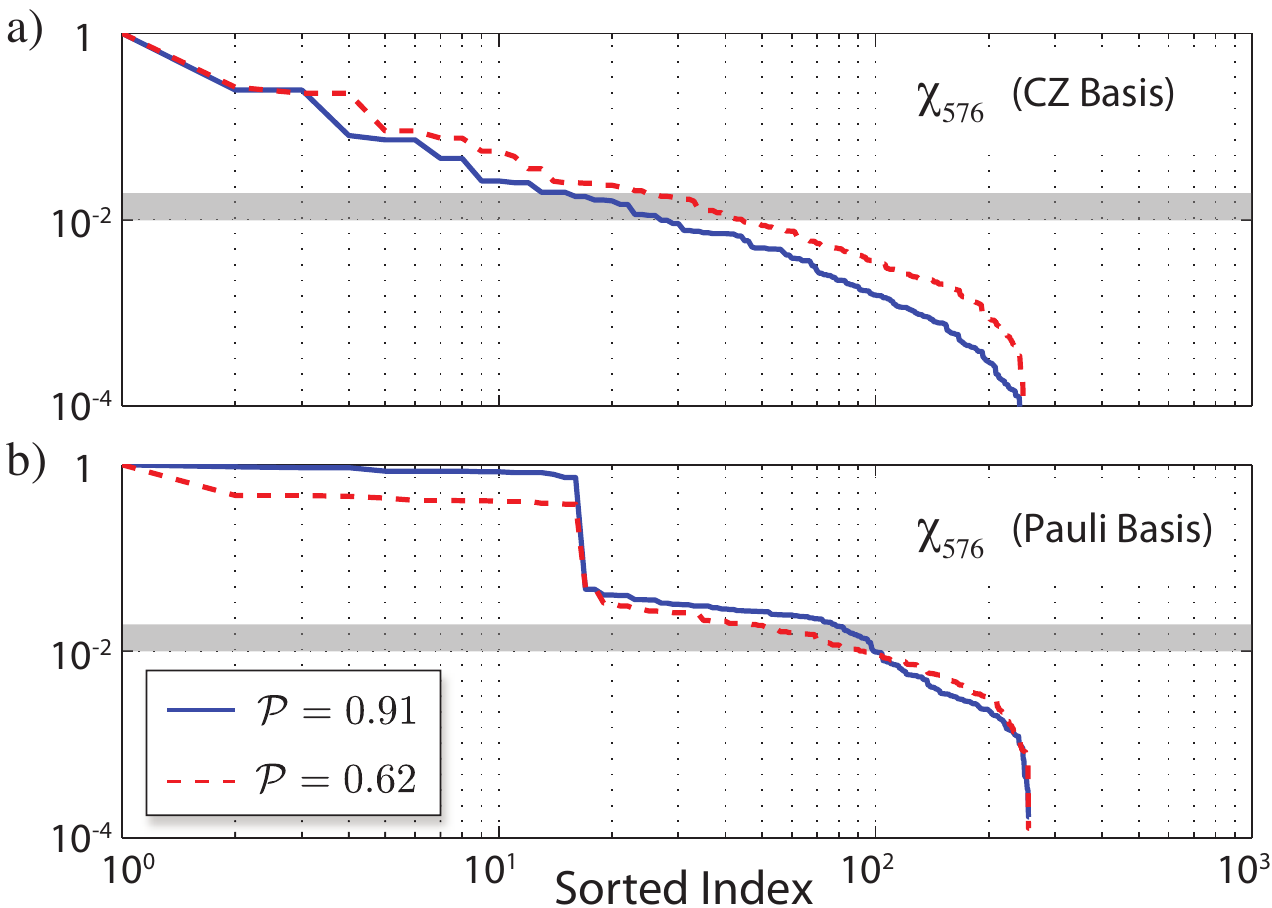,width=\columnwidth}
\caption{Absolute values of the 256 process matrix elements of
$\chi_{576}$ for our lowest and highest noise level, sorted by relative magnitude (with
respect to the 1,1-element) in the \textsc{cz} basis (top) and the Pauli basis (bottom). The error threshold, which indicates the required number of configurations, is shown in grey.}
\label{fig:loglog}
\efig

Applying CQPT to larger systems will require careful attention to classical post-processing which---as in QPT---scales exponentially. The standard software we used here (see Appendix \ref{app:software}), can easily handle 2 and 3 qubit CQPT systems. For larger systems, more specialized software can increase speed by orders of magnitude, \eg, \cite{cosamp:08}.

A number of research directions arise from this work: incorporating knowledge of model structure properties; tightening the bounds on scaling laws; understanding how near-sparsity $s$ and rank $r$ vary with system dimension, $d$; pursuing highly efficient convex-computational algorithms;  and selection of optimal configurations. Compressive tomography techniques can also be applied to quantum metrology and Hamiltonian parameter estimation: for example, estimating selective properties of biological or chemical interest in molecular systems and nanostructures with typically sparse Hamiltonians \cite{yuen2010qpt}.

\appendix
\section{Norms}
Definitions of the norms used throughout the paper. For a vector $x\in\Cbf^n$,
\beq[eq:vecnorms]
\bea{rcl}
\normtwo{x} &=& \sqrt{x^\dagger x}=\sqrt{\sum_{i=1}^n |x_i|^2}
\\ 
\normone{x} &=& \sum_{i=1}^n |x_i|.
\eea
\eeq
For a matrix $A\in\Cbf^{m\times n}$ with $\rank(A)=r\leq\min\{m,n\}$
and singular values $\sigma_1\geq\sigma_2\geq\cdots\geq\sigma_r>0$,
\beq[eq:matnorms]
\bea{llcl} 
\mbox{Induced $\ell_2$ norm} & \norm{A}_2 
&=& 
\sup_{\normtwo{x}=1}\normtwo{Ax}
= 
\sigma_1
\\
\mbox{Frobenius norm} 
&
\norm{A}_{\rm fro} &=& \sqrt{\trace(A^{\dagger}A)}
=\sqrt{\sum_{i=1}^r \sigma_i^2}
\\
\mbox{Nuclear norm} 
&
\norm{A}_\star &=& \trace(\sqrt{A^{\dagger}A})
=\sum_{i=1}^r \sigma_i
\eea
\eeq
%


\section{Restricted isometry property from a concentration inequality} 
\label{app:rip}
A common approach to establish the restricted isometry property (RIP),
Eqn.(3) in the paper, for a matrix $A\in\Cbf^{m\times n}$ with $m<n$
is by introducing randomness in the elements of this matrix. This
approach benefits from measure concentration properties of random
matrices.  For QPT for the measurment matrix $\Phi\in\Cbf^{m\times
d^4}$ in Eqn.(2) of the paper, we show how to achieve this with random
preparation of the intial states and a random selection of the
measurement operators. The proof is based on the results in
\cite{Baraniuk:2008} which show that if $\Phi$ is a random matrix
which satisfies the concentration property,
\beq[eq:concent]
\prob{ | \normtwo{\Phi x}^{2}-\normtwo{x}^{2} | 
\geq
\delta_s \normtwo{x}^{2} } 
\leq 2e^{-2mC_3(\delta_s)},
\eeq
for all $x\in\Cbf^{d^4}$, where $\delta_s\in(0,1)$ and $C_3(\delta_s)$ only
depends on $\delta_s$, then $\Phi$ satisfies the RIP,
\beq[eq:rip1]
(1-\delta_s)\normtwo{x_s}^2
\leq
\normtwo{\Phi x_s}^2
\leq
(1+\delta_s)\normtwo{x_s}^2
\eeq
for all $s$-sparse $x_s\in\Cbf^{d^4}$. This version of RIP is equivalent
to 
\refeq{rip}.

In classical signal processing, each element of the $\Phi$ matrix can
be independently selected from a random distribution such as Gaussian
or Bernoulli. For QPT there is no freedom for random independent
selection of every element of the $\Phi$ matrix.  However, as
described in the paper, the \emph{rows} of $\Phi$ can be independently
and randomly selected. To see this, recall that for each experimental
configuration we can initialize the system randomly in a state
$\rho\in\{\rho_i\in\Cbf^{d\times d}\}_ {i=1}^k$ and then measure an
observable $M$ randomly chosen from $\{M_j\in\Cbf^{d\times
d}\}_{j=1}^\ell$. The corresponding matrix $\Phi$ then has $m=k\ell$
independent random rows $\{\phi_i^\dag\in\Cbf^{1\times N}\}_{i=1}^m$
with correlated elements of each row since they are functions of the
same $M$ and $\rho$. Observe, however, that although $\Phi$ is a
random matrix, because it is constructed from quantum states and
observables of a finite dimensional system, it is bounded.  As a
consequence, $\forall x\in\Cbf^{d^4}$, we get,
\beq[eq:bound]
\bea{rcccl}
(w_\ell/m)\normtwo{x}^2 
&\leq& 
x^\dag(\phi_i\phi_i^\dag)x &\leq& (w_u/m)\normtwo{x}^2
\\
\ell\normtwo{x}^2 &\leq& \Ebf\normtwo{Ax}^2 &\leq& u\normtwo{x}^2
\eea
\eeq
where $\Ebf$ denotes expectation with respect to $\Phi$ and
$w_u,w_\ell,u,\ell$ are constants.  Next we apply,

\textbf{Hoeffding's concentration inequality}
\begin{em}
Let $v_{1},...,v_{m}$ be independent bounded random variables such
that $v_{i}$ falls in the interval $[a_{i},b_{i}]$ with probability
one. Then for $S=\sum_{i}v_{i}$ and any $t > 0$ we have,
\beq[eq:hoef]
\bea{rcl}
\prob{S-{\bf E}(S)\geq t}
&\leq& 
e^{-2t^{2}/\sum_{i}(b_{i}-a_{i})^{2}}
\\
\prob{S-{\bf E}(S)\leq -t}
&\leq& 
e^{-2t^{2}/\sum_{i}(b_{i}-a_{i})^{2}}
\eea
\eeq
\end{em}
\noindent In our problem $v_{i}=|\phi_{i}^\dag x|^{2}$ and
$S=\normtwo{\Phi x}^{2}$. From the above inequalities and the
relations in \refeq{bound} we find $\forall t_+, t_->0$ and $\forall
x$,
\beq[eq:hoef1]
\bea{rcl}
\prob{S-u\normtwo{x}^{2} \geq t_+} 
&\leq& 
\prob{S-{\bf E}(S) \geq t_+}
\\
&\leq& e^{-2t_+^{2}/(w_{u}-w_{\ell})^{2}}
\\
\prob{S-l\normtwo{x}^{2} \leq -t_-} 
&\leq& 
\prob{S-{\bf E}(S)\leq -t_-}
\\
&\leq& 
e^{-2t_-^{2}/(w_{u}-w_{\ell})^{2}}
\eea
\eeq
%
%
The choice of $t_+=(\delta_s+1-u)\normtwo{x}^{2}$ and
$t_-=(\ell-1+\delta_s)\normtwo{x}^{2}$ in the above inequalities
yields
\beq[eq:concent1]
\prob{ | \normtwo{\Phi x}^{2}-\norm{x}^{2}_{l_{2}^{m}} |
\geq \delta_s \normtwo{x}^{2} }
\leq 2e^{-2m(\delta_s+\epsilon)^{2}/(w_u-w_\ell)^{2}}
\eeq
with $\epsilon=\min\{1-u,\ell-1\}$.  We also need $t_+$ and $t_-$ to
be positive that imposes the condition $1-\delta_s<\ell\leq
u<1+\delta_s$. Since the obervable $M$ can be scaled by any real
factor, a sufficient condition is $u/\ell<(1+\delta_s)/(1-\delta_s)$.

Next we reproduce the connection between the measure concentration
\refeq{concent1} and restricted isometry as demonstrated in
\cite{Baraniuk:2008}: Let $X_s$ be a set of vectors with cardinality
$s$: $\#(X_s)=s$. We choose a set $Y\subset X_s$ such that
$\normtwo{y}=1$ for all $y\in Y$, we have $\min_{y\in
Y}\normtwo{x-y}\leq \delta_s/4$ for all $x\in X_s$.  The cardinality
of such a set $Y$ can always be chosen to be smaller than
$(12/\delta_s)^s$ \cite{Lorentz}.  There from \refeq{concent1} we find
\[
\prob{ | \normtwo{\Phi y}^{2}-1 |
\geq \delta_s/2} 
\leq 2(12/\delta_s)^se^\frac{-2m(\delta_s/2+\epsilon)^{2}}{(w_u-w_\ell)^{2}}
\]
or equivalently 
$
 1-\delta_s/2 \leq \normtwo{\Phi y}^{2} \leq 1+\delta_s/2
$
holds with probability exceeding
\[
P=1-2(12/\delta_s)^s \exp(-2m(\delta_s/2+\epsilon)^{2}/(w_u-w_\ell)^{2}).
\]
\noindent Define z to be the smallest number such that $\normtwo{\Phi
x'}\leq 1+z$ for all $x'$ with $\normtwo{x'}=1$.  For a vector $y\in
Y$ we have,
\[
\normtwo{\Phi x'}\leq \normtwo{\Phi y}
+\normtwo{\Phi (x'-y)}\leq 1+\frac{\delta_s}{2}+(1+z)\frac{\delta_s}{4}
\]
from which it follows that $z<\delta_s$, for any $0<\delta_s<1$. In a
similar fashion we can prove $1-\delta_s \leq \normtwo{\Phi x'}$.
This completes the proof that RIP \refeq{rip} holds with probability
exceeding $P$ for all $x\in X_s$. The number of sets $X_s$ with $\#
X_s=s$ is ${{N}\choose{s}}\leq (eN/s)^s$. Therefore RIP fails to be
satisfied with probability
$2\exp(-2m(\delta_s/2+\epsilon)^{2}/(w_u-w_\ell)^{2}
+s[log(eN/s)+log(12/\delta_s)])$.  For a sufficiently small constant
$C_0$, if $C_0s\leq m/log(N/s)$, we can find a constant $0<C_3$ such
that the probability of a failure of RIP becomes smaller than
$\exp(-C_3m)$ provided that $C_3\leq
2m(\delta_s/2+\epsilon)^{2}/(w_u-w_\ell)^{2}-s[log(eN/s)+log(12/\delta_s)]$.
This guaranteed exponentially small chance of RIP failure is the key to the
logarithmic scaling of the resources in CQPT.
If RIP is satisfied the $l_1$ norm minimization algorithm works to find a sparse solution.
Here we proved that by increasing the number of configurations $m$ would exponentially 
decrease the chance of RIP failure.
This completes the connection between the concentration measure
\refeq{concent1} and the restricted isometry property.

\section{Performance of the algorithm}

\noindent In Ref. \cite{Candes:08}, the accuracy of the $\ell_1$-norm
minimization problem is given by \refeq{perf}. 
The parameters $C_1$ and $C_2$ are explicitly given in terms of the
isometry constant $\delta_s$:
\beq[eq:c12]
C_1=\frac{2+(2\sqrt{2}-2)\delta_s}{1-(\sqrt{2}+1)\delta_s},
\quad
C_2=\frac{4\sqrt{1+\delta_s}}{1-(\sqrt{2}+1)\delta_s} 
\eeq

To present all the distances based on $l_1$-norm we can use
$||y||_{l_1}\leq ||y||_{l_2}\leq \sqrt{D} ||y||_{l_1}$, for a 
$D$-dimensional vector $y$ and obtain the algorithm performance as
\beq[eq:perf]
\normone{\vec\chi^\star-\vec\chi_0}
\leq
\frac{C_1d^2}{\sqrt{s}}
\normone{\vec\chi_0(s)-\vec\chi_0}+d^2C_2\varepsilon
\eeq
However the performance inequality presented in the paper has a tighter bound. 
\section{Sparsity assumption certification}

\noindent A test to certify the sparsity assumption can be concluded
from \refeq{perf} and the probability of RIP being satisfied exceeding
$1-e^{-mC_3(\delta_s)}$ for $m$ configurations. Suppose an estimate
$\chi_m$ is obtained for $m$ configurations. If the measure
$\normone{\chi_{m+1}-\chi_m}$, which quantifies an incremental
improvement in the estimated process matrix, converges toward zero for
a polynomially large $m$, the sparsity assumption is certified.

\section{Normalization and Precision Issues}
\label{app:normalization}
\noindent In the formulation of CQPT a random selection of the
expectation values $y_{M_i,\rho_i}$ are not available in our
experiment. Due to photon loss the detector counts are not conclusive,
hence, a complete set of counts corresponding to a complete set of
observables is required to produce meaningful expectation values
$y_{M_i,\rho_i}$. A solution to this problem is to limit the
measurements to few-body observables. For $k$-body measurements a
total number of $2^k$ complementary observables need to be
measured. Since $m$, the number of measurements, is exponentially
small we can choose $k$ limited to few-body operators, $k=k_{max}$,
and even single-body as we did in the experiment. For a fully random
selection of observables, the total number of measurements $m$ will be
increased by a constant factor $2^k_{max}$. Still this number is
exponentially small. This redundancy, however, can be avoided by using
the outcomes of all $2^k$ observables. This selection scheme is not
fully random, rather it is a deterministic-random way of choosing
observables.

\begin{small}
\begin{table}[t]
\centering
\begin{tabular}{c|c|c|c|c}

Inputs & Observables & $m$ & ${\cal F}(\chi_{576},\chi_{m})$ 
& ${\cal F}(U_{CZ},\chi_m)$ 
\\
\hline
\hline
\stak{HVDR} & \stak{HVDARL} & 576 & 1 & 0.88\\
\hline
\stak{HVDR} & \{RI,IR\} & 32 & 0.98 & 0.89\\
  & \{DI,ID\} &  & 0.97 & 0.87\\
\hline
\stak{HVDR} & \stak{RL} & 64 & 0.95 & 0.86\\
  & DAxDA & & 0.95 & 0.86\\
\hline
\stak{VDR} & \{RI,IR\} & 18 & 0.94 & 0.86\\
  & \{DI,ID\} & & 0.93 & 0.88\\
\hline
\stak{VDR} & \stak{RL} & 36 & 0.94 & 0.87\\
        & \stak{DA} &    & 0.94 & 0.84\\
\end{tabular}
\caption{Fidelity assessment of some selected configurations that are
available in our experiment.}
\label{tab:cqpt}
\end{table}
\end{small}

As discussed in the paper, random selections of probabilities from the
full data set, although exhibiting results which are entirely
consistent with compressive sensing theory, are inconsistent with how
data is actually collected in this kind of standard photonic
experiment. In practice we are limited to measure few-body
observables.  For low $m$, the configurations must allow for
normalisation, i.e. we are restricted in how random our low-number
selections can be.  A selection of some of these permissible
configurations are shown in Table~\ref{tab:cqpt}. Here we see some of
the remarkable results promised by the theory of compressed sensing,
\eg, a 98\% fidelity from 32 configurations and a 94\% fidelity from
only 18 configurations.

Another issue to consider is experimental precision. The expectation
values of $k$-body observables of random states reduce for a larger
$k$. This implies the need for a larger number of statistical
samples. Fortunately, this issue is not a problem for our scheme since
we can take $k$ as small as we want, as discussed above.

\section{Classical postprocessing}
\label{app:software}
\noindent The estimation results computed from the experimental data
were all obtained by solving equation 2 in the main text by using
``off-the-shelf'' MATLAB based software. Specifically, we used
\textit{YALMIP} to call the convex solver \textit{SDPT3}
\cite{yalmip:04,Sdpt3}. On a standard desktop it takes about 2 sec of
CPU-time to solve (2) for the full 576 configuration set. This
software can handle 3-qubit systems but it is more advisable to
migrate to more specialized software where orders of magnitude speed
increases are possible, \eg, \cite{cosamp:08}.

\section{Experimental Details}
\label{app:experiment}
\begin{figure}[t]
\includegraphics[width=\columnwidth, angle=0]{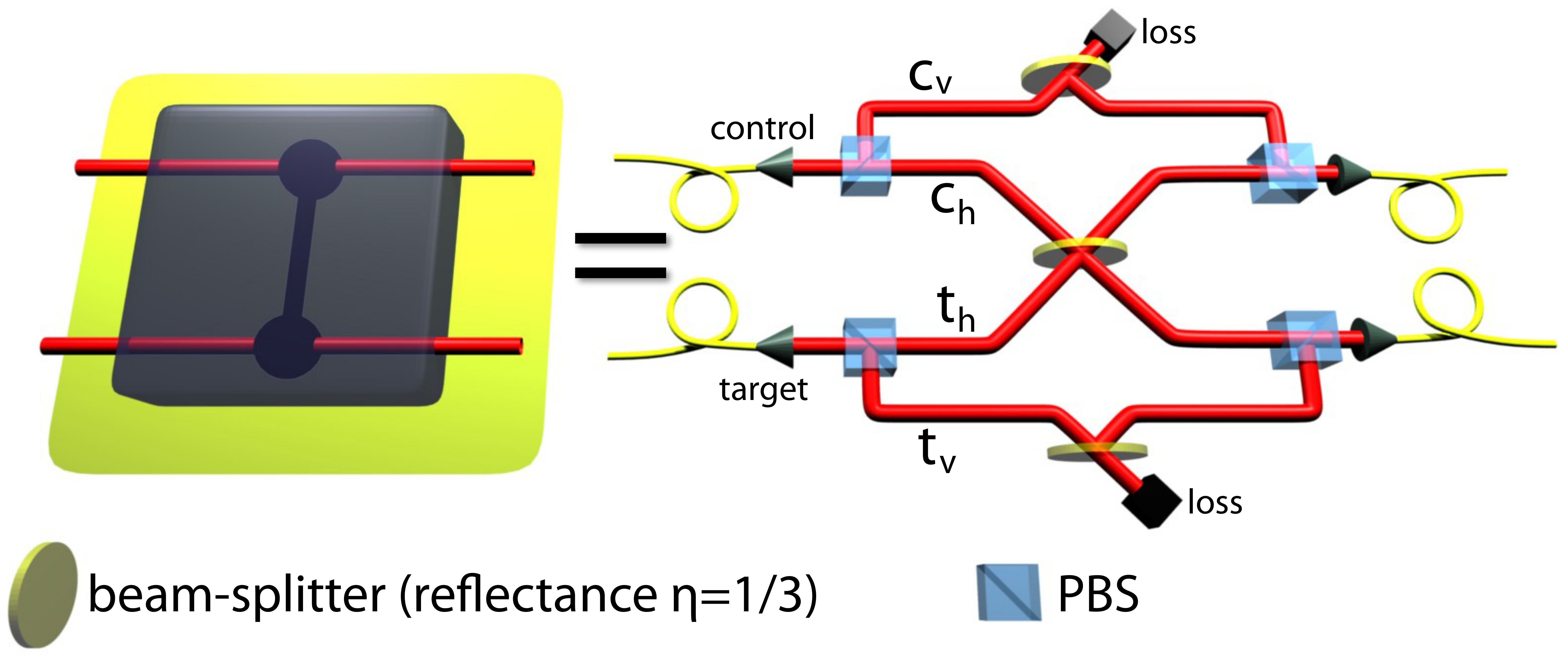}
\caption{Detailed representation of the \textsc{cz}-gate in dual rail
encoding. Each qubit is represented by two paths, one for each logical
basis state, $\ket{0}=\ket{H}$ and $\ket{1}=\ket{V}$
\cite{langford2005dse,okamoto2005doq,kiesel2005loc}.}
\label{fig:czdetail}
\end{figure}

\noindent The quantum gate used in the experiment is a photonic
controlled-phase gate, Fig.~\ref{fig:czdetail}
\cite{langford2005dse,okamoto2005doq,kiesel2005loc}. It is based on a single partially polarising
beam splitter (PPBS), having different reflectivity,
$\eta_{V}{=}\frac{1}{3}$, $\eta_{H}{=}0$, for the horizontal and the
vertical polarisation of input photons. Due to two-photon
interference, the input state $\ket{VV}$ undergoes a $\pi$ phase shift
$\ket{VV}\rightarrow-\ket{VV}$ whenever the two photons leave the PPBS
through different output ports. Correct operation of the gate is
signalled by a coincidence detection in these output modes; the gate
is thus probabilistic, with a success probability of $1/9$.

The gate acts on photonic qubits created via spontaneous parametric
downconversion (SPDC). Downconversion is intrinsically a random
process: consequently the created states contain small amounts of
higher-order emission---e.g. $\ket{22}$ as well as the desired
$\ket{11}$---which appear as decoherence in a quantum process
\cite{weinhold2008upq,barbieri2009pdo}. The ratio of higher order
terms to the desired photon pair number increases with the pair
creation probability, which in turn is proportional to the pump laser
power. Once can therefore---to some extent---control the decoherence
in a process via the laser power.

In order to cover a comprehensive range of decoherence, we performed
six experiments with 2-photon states directly created via a single
SPDC emission, and one experiment with 4-photon states created in two
independent SPDC sources, where one photon of each SPDC process was
used as a trigger. The latter experiment is more representative of
large-scale systems, where independent photon sources will be
required. It has significantly reduced count rates, and reduced
two-photon interference between photons in the quantum gate due to
both the pump-induced decoherence and group-velocity mismatch
\cite{weinhold2008upq}, reflected in the low purity of the process in
this case of 0.62.

Typical count rates for 2-photon experiments are $2000$ coincident
counts per second, full QPT, building up reasonable statistics, takes
about $2.5$ hours; in contrast, 4-photon experiments have much lower
rates, $1$ four-fold coincidence per second, and take $2$
\emph{days}. The 32-configuration CQPT reduces tomography times to 8
minutes and 2.6 hours respectively: a clear advantage.

\begin{figure}[t]
\centering
\epsfig{file=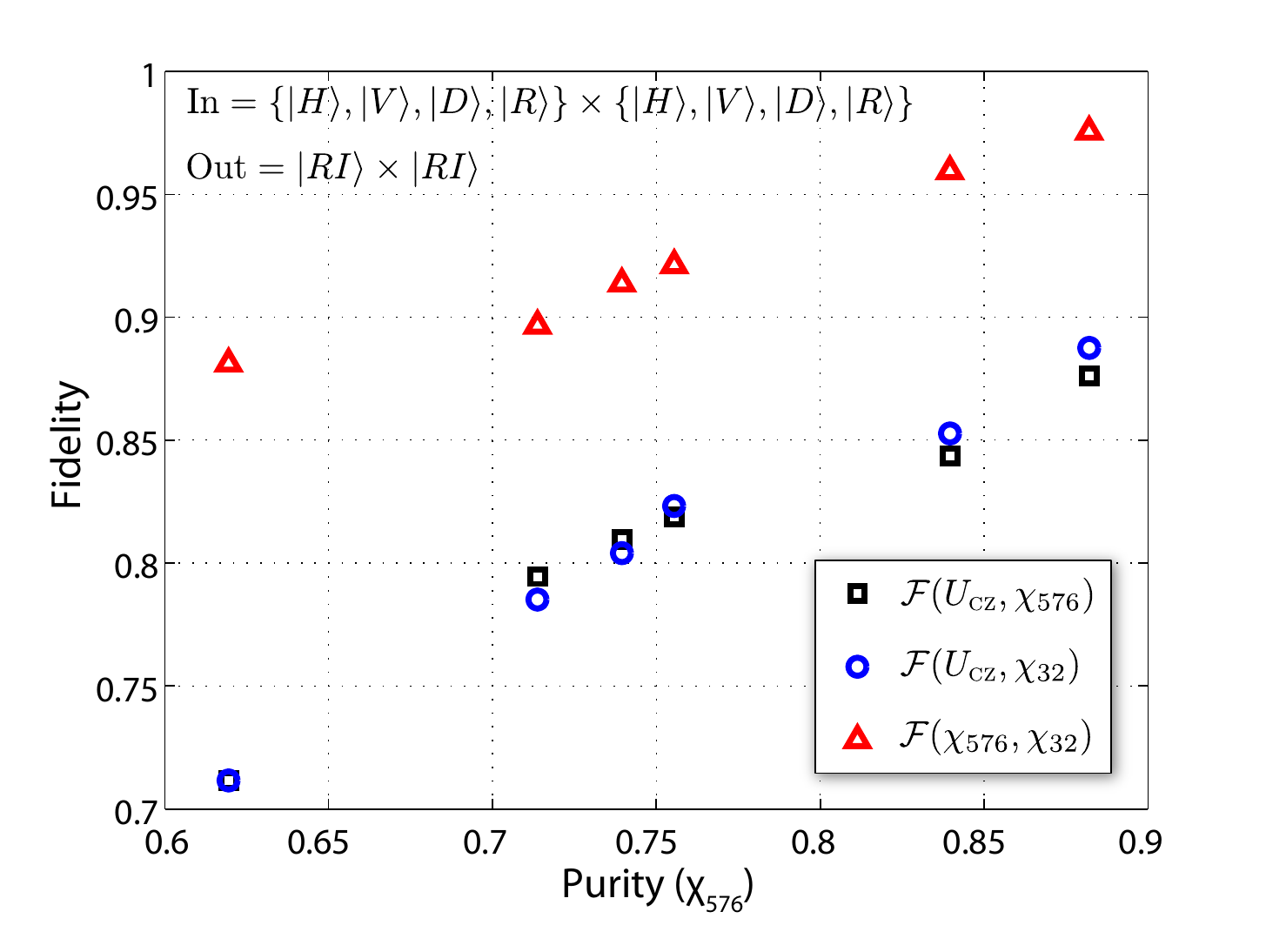,width=\columnwidth}
\caption{Fidelities vs. purities for $m=32$ corresponding to the configurations in Table \ref{tab:cqpt}.}
\label{fig:purity}
\end{figure}

Fig.~\ref{fig:purity} shows the effect of varying laser pump power on
CQPT estimation accuracy for one of the single-observable
configurations from Table \ref{tab:cqpt}. Specifically for the 32
configurations arising from all combinations of the 16 inputs
\stk{HVDR} and 2 outcomes \{RI,IR\}. As pump power increases, the
process purity, as measured by $\trace(\chi_{576}^2)/16$ decreases; effectively the signal to noise ratio deteriorates. 
As might be expected, the worst-case fidelity decreases with process
purity. The estimated channel fidelity is however remarkably robust,
staying very close to the actual channel fidelity. 


\section{Simulation results}
\label{app:sim2qb}
QPT is performed by solving (\ref{eq:dec}) with noise-free experiments ($\epsilon=0$) for a system
designed to be a 2-qubit quantum Fourier transform (QFT) with
unitary representation $U_\qft\in\mathbb{C}^{4\times 4}$, which
interacts with an unknown environment via the total \emph{constant}
Hamiltonian, $H{=}I_{e}\otimes H_\qft{+}\gamma\tilde{H}$ with
$\|\tilde{H}\|{=}1$; $\gamma$ is thus the interaction magnitude. The
simulated system $\chi_\textrm{sim}\in\mathbb{C}^{16\times 16}$ is extracted via the partial
trace over the environment for $\gamma\in\{0.5,1.0,1.25\}$. Each of
these induces a fidelity with respect to the ideal
unitary, $\mathcal{F}(U_\qft,\chi_\textrm{sim})\in\{0.70,0.80,0.95\}$
The estimates from (\ref{eq:dec}) are obtained in the singular value
decomposition (SVD)-basis [$\Gam_\alpha$ in (\ref{eq:osr})] of the ideal
QFT. The process matrix of the ideal unitary in this basis is
maximally sparse with the single non-zero 1,1-element equal to $n=4$
\cite{Kosut:08qpt}. The environmental interactions make the process
matrix almost sparse as defined in (\ref{eq:dec}). 

To form the measurement matrix $\Phi\in\mathbb{C}^{m\times 256}$, we
randomly generated 4 and 16 input pairs
$\ket{\psi_1}\otimes\ket{\psi_2}$ and 2, 4, and 6 random selections
from the \emph{single-body} Pauli observables
$\{IX,IY,IZ,XI,YI,ZI\}$. This gives 6 configurations with
$m\in\{8,16,24,32,64,96\}$, for which $u/\ell\approx 1.3$ ensuring
$\delta\approx 0.13$.  Fig.~\ref{fig:qft2} shows the fidelities $\mathcal{F}(\chi_{m},\chi_\textrm{sim})$ of the reconstructed estimates $\chi_{m}$ and the simulated process matrices $\chi_\textrm{sim}$ for all
18 combinations of $m$ and interaction magnitudes $\gamma$.

\bfig \btab{c}
\epsfig{file=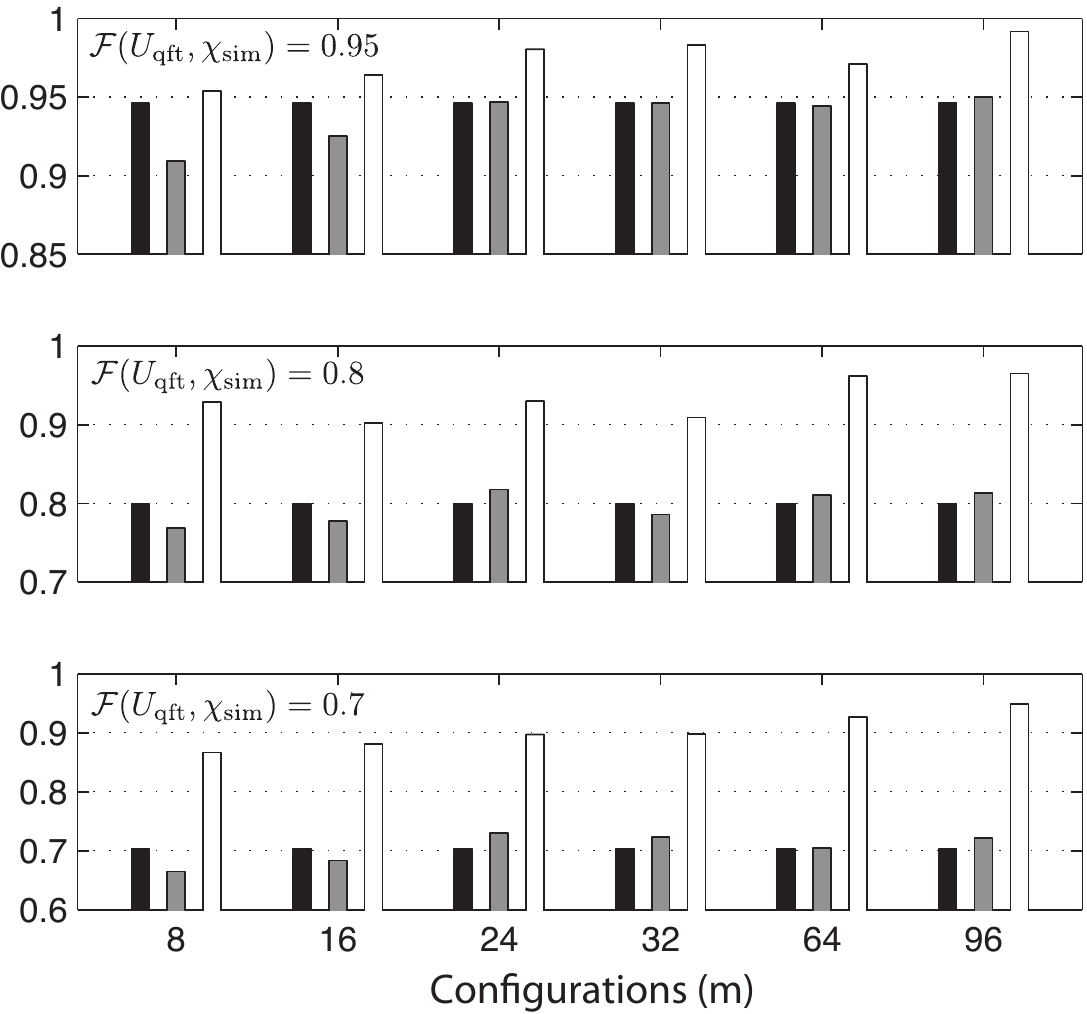,width=0.95\columnwidth}
\etab
\caption{Fidelities vs. configurations for each process matrix estimate
$\chi_{m}$ from (\ref{eq:dec}) in the SVD basis of the ideal QFT unitary.  Black bars: simulated compared to ideal process $\mathcal{F}(U_\qft,\chi_\textrm{sim})$. Gray bars: estimate compared to ideal
$\mathcal{F}(U_\qft,\chi_{m})$. White bars: estimate compared to simulated process $\mathcal{F}(\chi_{m},\chi_\textrm{sim})$.}
\label{fig:qft2}
\efig

These results arise from the relative sparsity of the process matrix
in the SVD-basis of the ideal QFT. Fig.~\ref{fig:qft1} shows 3D bar
plots of the real and imaginary elements of the true and estimated
process matrices for $m=64$, $\fchn(U_\qft,\chi_{\rm sim})=0.70$, and
$\fchn(\chi_{64},\chi_{\rm sim})=0.93$. In the SVD-basis (row 2), the
true process matrix exhibits the expected large 1,1-element with the
remaining elements much smaller by comparison. The estimated channel
fidelity is 0.71.

\bfig
\epsfig{file=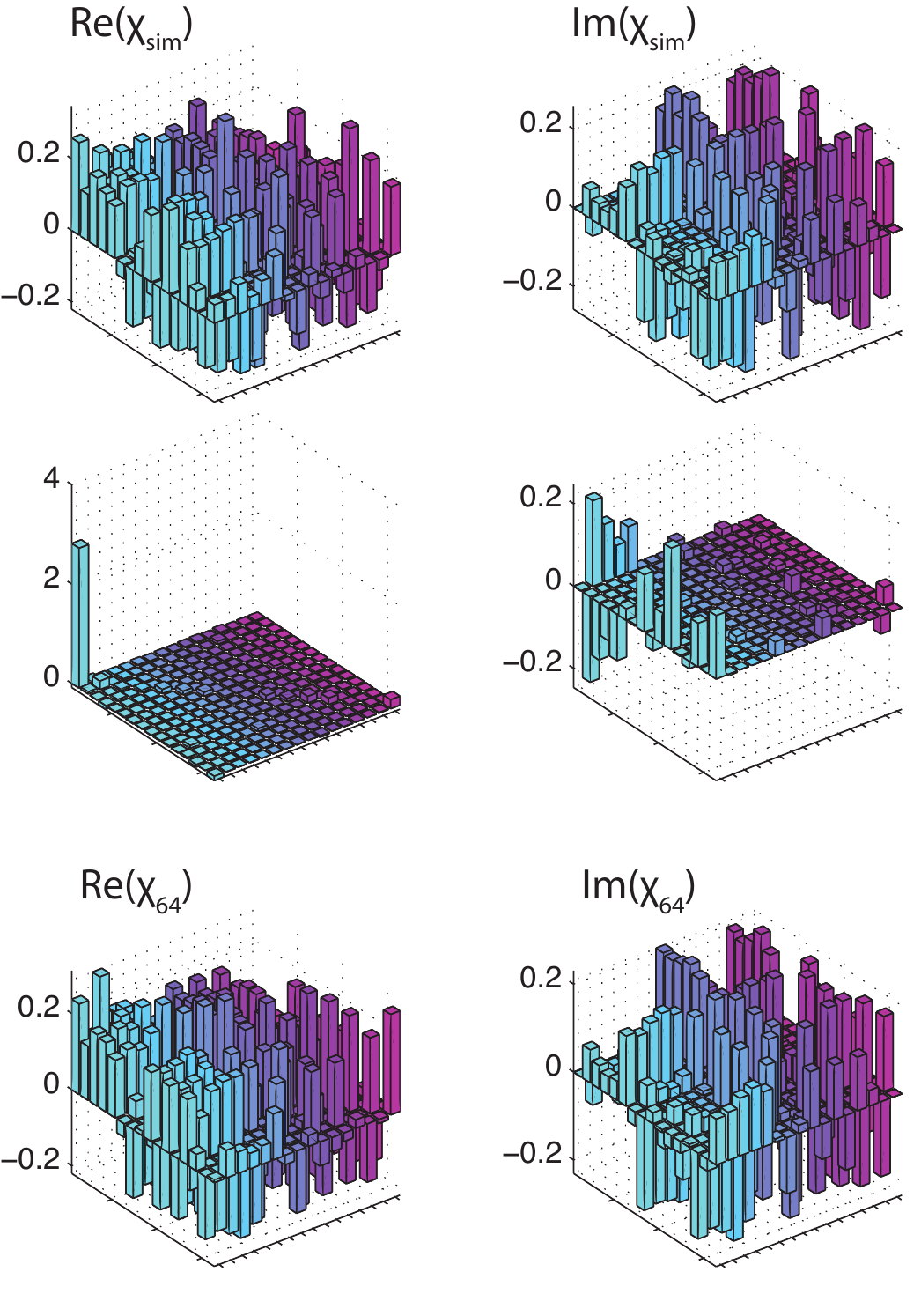,width=.8\columnwidth}
\caption{ Real and imaginary $\chi$ elements for $m{=}64$, $\mathcal{F}(U\textrm{qft},\chi_\textrm{sim})=0.71$, $\gamma{=}1.25$. Row 1: True process matrix in the natural basis.  Row
2: True process matrix in SVD-basis of ideal unitary.  Row 3:
Estimated process matrix projected to the natural basis
, $\mathcal{F}(U_\textrm{qft},\chi_{m})=0.71$.}
\label{fig:qft1}
\efig

\bfig[h]
\btab{c}
\epsfig{file=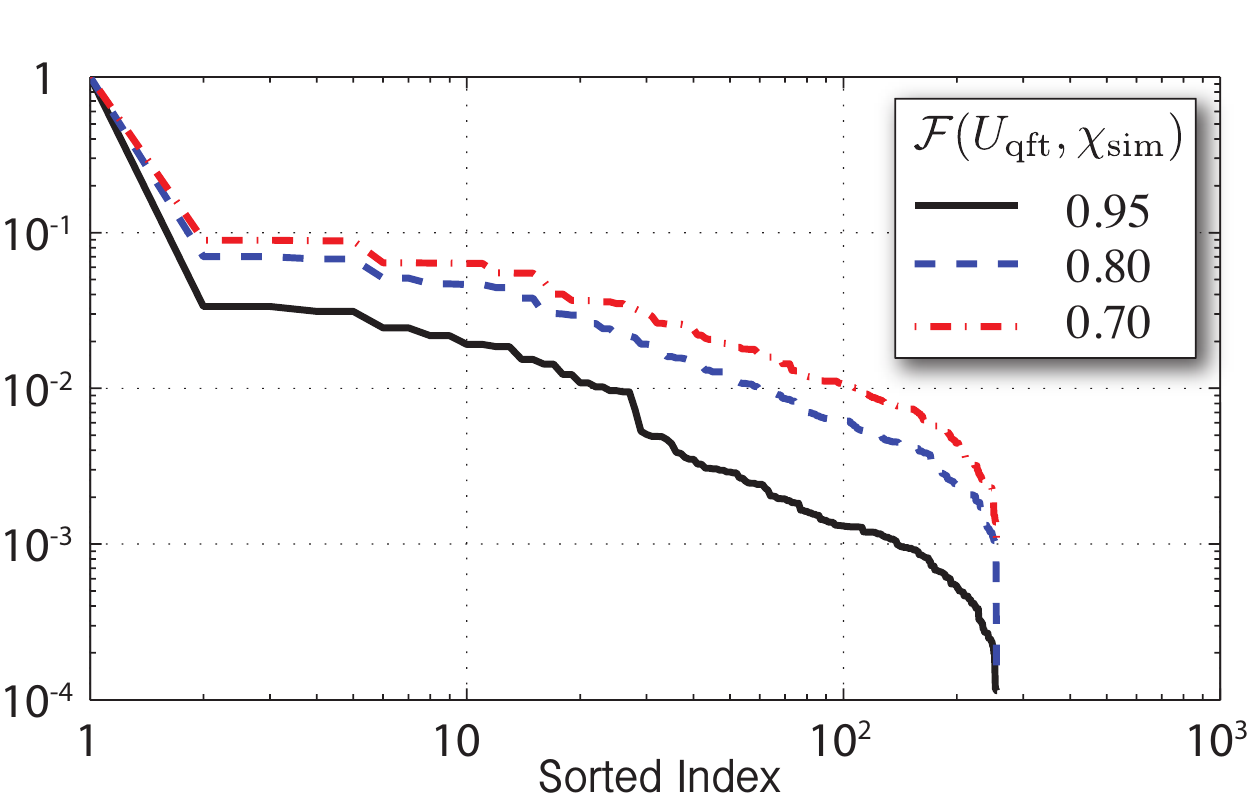,width=.8\columnwidth}
\etab
\caption{Absolute values of the 256 process matrix elements of
$\chi_\svd^\true\in\Cbf^{16\times 16}$ sorted by relative magnitude
(with respect to the 11-element) for 
$\fchn(U_\qft,\Scal^\true)\in\{0.95,0.80,0.70\}$.
}
\label{fig:loglog1}
\efig

In Fig.~\ref{fig:qft2}, $\fchn(\chi_m,\chi_{\rm sim})$ (white bars)
trends to increase with $m$, more so for $\fchn=0.7$ than for
$\fchn=0.95$, and rises a bit sharply at different $m$ values. Just as for the experimental results, this
can be connected to the actual sparsity of the simulated process matrices. Figure~\ref{fig:loglog1}, just like Fig. \ref{fig:loglog} in the main text, shows the absolute sorted process matrix elements relative to the 1,1-element. Where each plot crosses the threshold of $0.02$, we see that the number of
elements above this value increases with decreasing decoherence
$\gamma$. If these are taken as the $s$-sparse approximation levels
indicated in the theory (\ref{eq:dec}), then (approximately)
$s\in\{30,50,100\}$ correspond to
$\mathcal{F}(U_\qft,\chi_\textrm{sim})\in\{0.95,0.80.0.70\}$.  This
correlates well with how $\mathcal{F}(\chi_{m},\chi_\textrm{sim})$
varies with resources $m$.

\section{Beyond 2-qubits}
\label{app:simnqb}
\bfig[t!]
\btab{c}
\epsfig{file=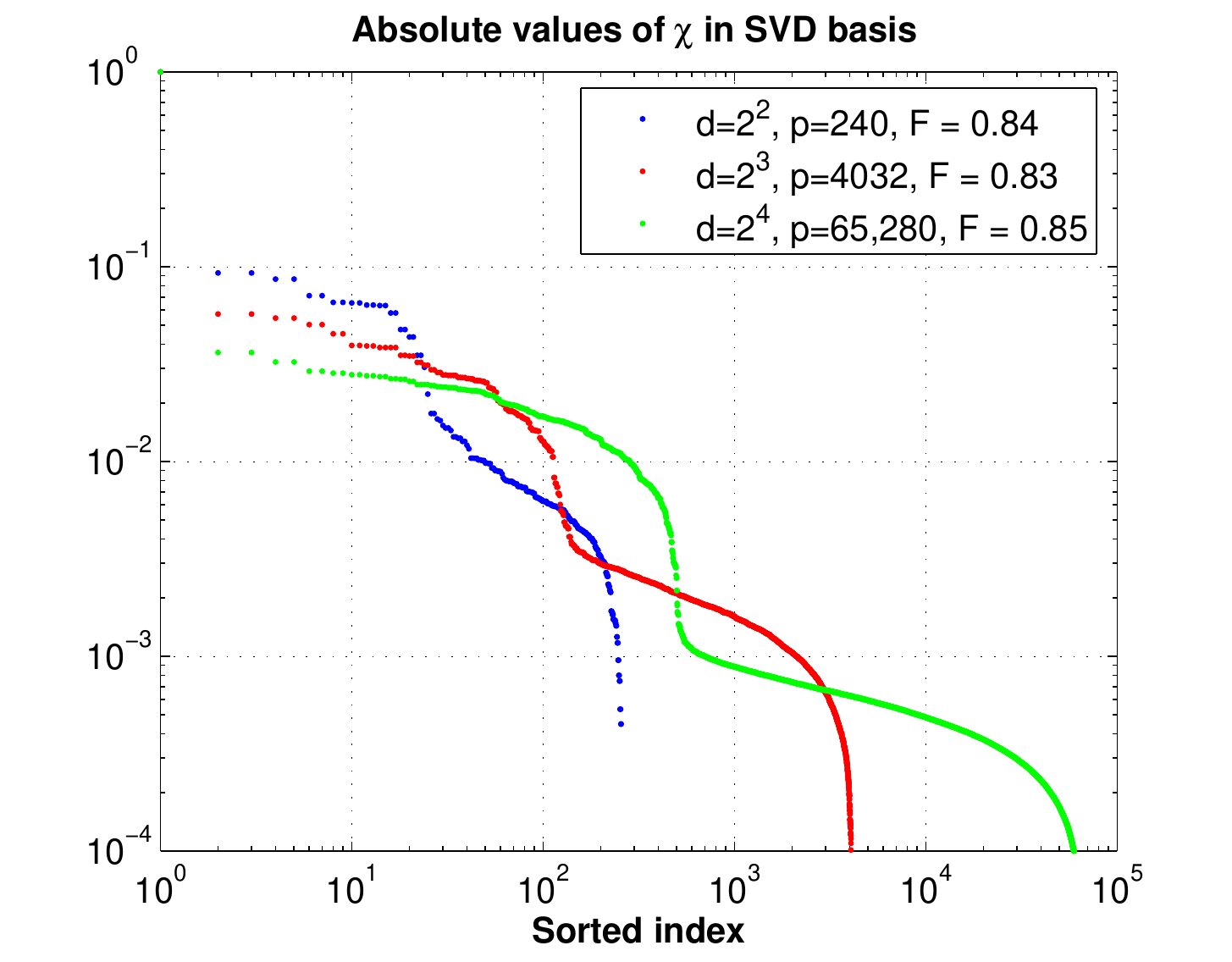,width=.9\columnwidth}
\etab
\caption{
Absolute values of the process matrix elements sorted by
relative magnitude (with respect to the 11-element) all in the ideal
SVD basis (in this case for an identity operator on the system) for
three cases: 
{\bf blue} $\chi_{\rm sim}\in\Cbf^{16\times 16}$ with $\fchn(I_4,\chi_{\rm
sim})=0.84$;
{\bf red} $\chi_{\rm sim}\in\Cbf^{64\times 64}$ with $\fchn(I_8,\chi_{\rm
sim})=0.83$;
{\bf green} $\chi_{\rm sim}\in\Cbf^{256\times 256}$ with
$\fchn(I_{16},\chi_{\rm sim})=0.85$;
}
\label{fig:loglog234}
\efig
Standard QPT scales exponentially, thus for 3 and 4 qubits the number
of required experimental configurations is, respectively 4,032 and
65,280. As we have shown theoretically, experimentaly, and lastly via
the previous simulations, CQPT shows quite a different scaling.
Fig.~\ref{fig:loglog234} shows the absolute values, sorted by relative
magnitude, of the process matrices arising from a random selection of
a perturbed system near identity, \ie, a quantum memory, corresponding
to similar fidelities. The process matrices elements are shown in a
basis corresponding to the ideal identity. Again taking 0.01 as a
threshold we see that for 2-qubits we get $m\approx 20$ which is
similar to our experimental results and those supported by the plots
in Figures \ref{fig:loglog} and
\ref{fig:loglog1}. Fig.~\ref{fig:loglog234} predicts for 3-qubits
$m\approx 100$, and for 4-qubits $m\approx 300$.  These simulation
results show first that the process matrices are compressible, and in
addition are consistent with the experimental results in
Fig.~\ref{fig:loglog}. To actually perform the estimaton, that is
solve \refeq{dec}, as previously mentioned, requires specialized
compressed sensing algorithms optimized for speed and efficiency, \eg,
\cite{cosamp:08}.


\begin{acknowledgments}
We thank J. Romberg and S. Jafarpour for discussions. We acknowledge funding by the ARC Discovery and Federation Fellow programs and an IARPA-funded US Army Research Office contract. RLK and HR are supported by DARPA Grant FA9550-09-1-0710 administered through AFOSR.
\end{acknowledgments}


\begin{thebibliography}{38}
\expandafter\ifx\csname natexlab\endcsname\relax\def\natexlab#1{#1}\fi
\expandafter\ifx\csname bibnamefont\endcsname\relax
  \def\bibnamefont#1{#1}\fi
\expandafter\ifx\csname bibfnamefont\endcsname\relax
  \def\bibfnamefont#1{#1}\fi
\expandafter\ifx\csname citenamefont\endcsname\relax
  \def\citenamefont#1{#1}\fi
\expandafter\ifx\csname url\endcsname\relax
  \def\url#1{\texttt{#1}}\fi
\expandafter\ifx\csname urlprefix\endcsname\relax\def\urlprefix{URL }\fi
\providecommand{\bibinfo}[2]{#2}
\providecommand{\eprint}[2][]{\url{#2}}

\bibitem[{\citenamefont{Nielsen and Chuang}(2000)}]{Nielsen:book}
\bibinfo{author}{\bibfnamefont{M.}~\bibnamefont{Nielsen}} \bibnamefont{and}
  \bibinfo{author}{\bibfnamefont{I.}~\bibnamefont{Chuang}},
  \emph{\bibinfo{title}{Quantum computation and quantum information}}
  (\bibinfo{publisher}{Cambridge University Press}, \bibinfo{year}{2000}).

\bibitem[{\citenamefont{Altepeter et~al.}(2003)\citenamefont{Altepeter,
  Branning, Jeffrey, Wei, Kwiat, Thew, O'Brien, Nielsen, and
  White}}]{altepeter2003aaq}
\bibinfo{author}{\bibfnamefont{J.~B.} \bibnamefont{Altepeter}},
  \bibinfo{author}{\bibfnamefont{D.}~\bibnamefont{Branning}},
  \bibinfo{author}{\bibfnamefont{E.}~\bibnamefont{Jeffrey}},
  \bibinfo{author}{\bibfnamefont{T.~C.} \bibnamefont{Wei}},
  \bibinfo{author}{\bibfnamefont{P.~G.} \bibnamefont{Kwiat}},
  \bibinfo{author}{\bibfnamefont{R.~T.} \bibnamefont{Thew}},
  \bibinfo{author}{\bibfnamefont{J.~L.} \bibnamefont{O'Brien}},
  \bibinfo{author}{\bibfnamefont{M.~A.} \bibnamefont{Nielsen}},
  \bibnamefont{and} \bibinfo{author}{\bibfnamefont{A.~G.} \bibnamefont{White}},
  \bibinfo{journal}{Phys. Rev. Lett.} \textbf{\bibinfo{volume}{90}},
  \bibinfo{pages}{193601} (\bibinfo{year}{2003}).

\bibitem[{\citenamefont{Mohseni and Lidar}(2006)}]{DQPT}
\bibinfo{author}{\bibfnamefont{M.}~\bibnamefont{Mohseni}} \bibnamefont{and}
  \bibinfo{author}{\bibfnamefont{D.}~\bibnamefont{Lidar}},
  \bibinfo{journal}{Phys. Rev. Lett.} \textbf{\bibinfo{volume}{97}},
  \bibinfo{pages}{170501} (\bibinfo{year}{2006}).

\bibitem[{\citenamefont{Lobino et~al.}(2008)\citenamefont{Lobino, Korystov,
  Kupchak, Figueroa, Sanders, and Lvovsky}}]{lobino2008ccq}
\bibinfo{author}{\bibfnamefont{M.}~\bibnamefont{Lobino}},
  \bibinfo{author}{\bibfnamefont{D.}~\bibnamefont{Korystov}},
  \bibinfo{author}{\bibfnamefont{C.}~\bibnamefont{Kupchak}},
  \bibinfo{author}{\bibfnamefont{E.}~\bibnamefont{Figueroa}},
  \bibinfo{author}{\bibfnamefont{B.~C.} \bibnamefont{Sanders}},
  \bibnamefont{and} \bibinfo{author}{\bibfnamefont{A.~I.}
  \bibnamefont{Lvovsky}}, \bibinfo{journal}{Science}
  \textbf{\bibinfo{volume}{322}}, \bibinfo{pages}{563} (\bibinfo{year}{2008}).

\bibitem[{\citenamefont{Mohseni et~al.}(2008)\citenamefont{Mohseni, Rezakhani,
  and Lidar}}]{Mohseni:08}
\bibinfo{author}{\bibfnamefont{M.}~\bibnamefont{Mohseni}},
  \bibinfo{author}{\bibfnamefont{A.~T.} \bibnamefont{Rezakhani}},
  \bibnamefont{and} \bibinfo{author}{\bibfnamefont{D.~A.} \bibnamefont{Lidar}},
  \bibinfo{journal}{Phys. Rev. A} \textbf{\bibinfo{volume}{77}},
  \bibinfo{pages}{032322} (\bibinfo{year}{2008}).

\bibitem[{\citenamefont{Emerson et~al.}(2007)\citenamefont{Emerson, Silva,
  Moussa, Ryan, Laforest, Baugh, Cory, and Laflamme}}]{emerson2007scn}
\bibinfo{author}{\bibfnamefont{J.}~\bibnamefont{Emerson}},
  \bibinfo{author}{\bibfnamefont{M.}~\bibnamefont{Silva}},
  \bibinfo{author}{\bibfnamefont{O.}~\bibnamefont{Moussa}},
  \bibinfo{author}{\bibfnamefont{C.}~\bibnamefont{Ryan}},
  \bibinfo{author}{\bibfnamefont{M.}~\bibnamefont{Laforest}},
  \bibinfo{author}{\bibfnamefont{J.}~\bibnamefont{Baugh}},
  \bibinfo{author}{\bibfnamefont{D.~G.} \bibnamefont{Cory}}, \bibnamefont{and}
  \bibinfo{author}{\bibfnamefont{R.}~\bibnamefont{Laflamme}},
  \bibinfo{journal}{Science} \textbf{\bibinfo{volume}{317}},
  \bibinfo{pages}{1893} (\bibinfo{year}{2007}).

\bibitem[{\citenamefont{Bendersky et~al.}(2008)\citenamefont{Bendersky,
  Pastawski, and Paz}}]{bendersky2008see}
\bibinfo{author}{\bibfnamefont{A.}~\bibnamefont{Bendersky}},
  \bibinfo{author}{\bibfnamefont{F.}~\bibnamefont{Pastawski}},
  \bibnamefont{and} \bibinfo{author}{\bibfnamefont{J.~P.} \bibnamefont{Paz}},
  \bibinfo{journal}{Phys. Rev. Lett.} \textbf{\bibinfo{volume}{100}},
  \bibinfo{pages}{190403} (\bibinfo{year}{2008}).

\bibitem[{\citenamefont{Branderhorst et~al.}(2009)\citenamefont{Branderhorst,
  Nunn, Walmsley, and Kosut}}]{branderhorst2009sqp}
\bibinfo{author}{\bibfnamefont{M.}~\bibnamefont{Branderhorst}},
  \bibinfo{author}{\bibfnamefont{J.}~\bibnamefont{Nunn}},
  \bibinfo{author}{\bibfnamefont{I.}~\bibnamefont{Walmsley}}, \bibnamefont{and}
  \bibinfo{author}{\bibfnamefont{R.}~\bibnamefont{Kosut}},
  \bibinfo{journal}{New Journal of Physics} \textbf{\bibinfo{volume}{11}},
  \bibinfo{pages}{115010} (\bibinfo{year}{2009}).

\bibitem[{\citenamefont{Haffner et~al.}(2005)\citenamefont{Haffner, Hansel,
  Roos, Benhelm, Chek-al kar, Chwalla, Korber, Rapol, Riebe, Schmidt
  et~al.}}]{haffner2005sme}
\bibinfo{author}{\bibfnamefont{H.}~\bibnamefont{Haffner}},
  \bibinfo{author}{\bibfnamefont{W.}~\bibnamefont{Hansel}},
  \bibinfo{author}{\bibfnamefont{C.}~\bibnamefont{Roos}},
  \bibinfo{author}{\bibfnamefont{J.}~\bibnamefont{Benhelm}},
  \bibinfo{author}{\bibfnamefont{D.}~\bibnamefont{Chek-al kar}},
  \bibinfo{author}{\bibfnamefont{M.}~\bibnamefont{Chwalla}},
  \bibinfo{author}{\bibfnamefont{T.}~\bibnamefont{Korber}},
  \bibinfo{author}{\bibfnamefont{U.}~\bibnamefont{Rapol}},
  \bibinfo{author}{\bibfnamefont{M.}~\bibnamefont{Riebe}},
  \bibinfo{author}{\bibfnamefont{P.}~\bibnamefont{Schmidt}},
  \bibnamefont{et~al.}, \bibinfo{journal}{Nature}
  \textbf{\bibinfo{volume}{438}}, \bibinfo{pages}{643} (\bibinfo{year}{2005}).

\bibitem[{\citenamefont{O'Brien et~al.}(2004)\citenamefont{O'Brien, Pryde,
  Gilchrist, James, Langford, Ralph, and White}}]{obrien2004qpt}
\bibinfo{author}{\bibfnamefont{J.}~\bibnamefont{O'Brien}},
  \bibinfo{author}{\bibfnamefont{G.}~\bibnamefont{Pryde}},
  \bibinfo{author}{\bibfnamefont{A.}~\bibnamefont{Gilchrist}},
  \bibinfo{author}{\bibfnamefont{D.}~\bibnamefont{James}},
  \bibinfo{author}{\bibfnamefont{N.}~\bibnamefont{Langford}},
  \bibinfo{author}{\bibfnamefont{T.}~\bibnamefont{Ralph}}, \bibnamefont{and}
  \bibinfo{author}{\bibfnamefont{A.}~\bibnamefont{White}},
  \bibinfo{journal}{Phys. Rev. Lett.} \textbf{\bibinfo{volume}{93}},
  \bibinfo{pages}{080502} (\bibinfo{year}{2004}).

\bibitem[{\citenamefont{Bialczak et~al.}(2010)\citenamefont{Bialczak, Ansmann,
  Hofheinz, Lucero, Neeley, O'Connell, Sank, Wang, Wenner, Steffen
  et~al.}}]{Bialczak}
\bibinfo{author}{\bibfnamefont{R.}~\bibnamefont{Bialczak}},
  \bibinfo{author}{\bibfnamefont{M.}~\bibnamefont{Ansmann}},
  \bibinfo{author}{\bibfnamefont{M.}~\bibnamefont{Hofheinz}},
  \bibinfo{author}{\bibfnamefont{E.}~\bibnamefont{Lucero}},
  \bibinfo{author}{\bibfnamefont{M.}~\bibnamefont{Neeley}},
  \bibinfo{author}{\bibfnamefont{A.}~\bibnamefont{O'Connell}},
  \bibinfo{author}{\bibfnamefont{D.}~\bibnamefont{Sank}},
  \bibinfo{author}{\bibfnamefont{H.}~\bibnamefont{Wang}},
  \bibinfo{author}{\bibfnamefont{J.}~\bibnamefont{Wenner}},
  \bibinfo{author}{\bibfnamefont{M.}~\bibnamefont{Steffen}},
  \bibnamefont{et~al.}, \bibinfo{journal}{Nature Physics}
  \textbf{\bibinfo{volume}{6}}, \bibinfo{pages}{409} (\bibinfo{year}{2010}).

\bibitem[{\citenamefont{Monz et~al.}(2009)\citenamefont{Monz, Kim, H\"{a}nsel,
  Riebe, Villar, Schindler, Chwalla, Hennrich, and Blatt}}]{monz2008rqt}
\bibinfo{author}{\bibfnamefont{T.}~\bibnamefont{Monz}},
  \bibinfo{author}{\bibfnamefont{K.}~\bibnamefont{Kim}},
  \bibinfo{author}{\bibfnamefont{W.}~\bibnamefont{H\"{a}nsel}},
  \bibinfo{author}{\bibfnamefont{M.}~\bibnamefont{Riebe}},
  \bibinfo{author}{\bibfnamefont{A.~S.} \bibnamefont{Villar}},
  \bibinfo{author}{\bibfnamefont{P.}~\bibnamefont{Schindler}},
  \bibinfo{author}{\bibfnamefont{M.}~\bibnamefont{Chwalla}},
  \bibinfo{author}{\bibfnamefont{M.}~\bibnamefont{Hennrich}}, \bibnamefont{and}
  \bibinfo{author}{\bibfnamefont{R.}~\bibnamefont{Blatt}},
  \bibinfo{journal}{Phys. Rev. Lett.} \textbf{\bibinfo{volume}{102}},
  \bibinfo{eid}{040501} (pages~\bibinfo{numpages}{4}) (\bibinfo{year}{2009}).

\bibitem[{\citenamefont{Childs et~al.}(2001)\citenamefont{Childs, Chuang, and
  Leung}}]{childs2001rqp}
\bibinfo{author}{\bibfnamefont{A.~M.} \bibnamefont{Childs}},
  \bibinfo{author}{\bibfnamefont{I.~L.} \bibnamefont{Chuang}},
  \bibnamefont{and} \bibinfo{author}{\bibfnamefont{D.~W.} \bibnamefont{Leung}},
  \bibinfo{journal}{Phys. Rev. A} \textbf{\bibinfo{volume}{64}},
  \bibinfo{pages}{012314} (\bibinfo{year}{2001}).

\bibitem[{\citenamefont{Boulant et~al.}(2003)\citenamefont{Boulant, Havel,
  Pravia, and Cory}}]{boulant2003rme}
\bibinfo{author}{\bibfnamefont{N.}~\bibnamefont{Boulant}},
  \bibinfo{author}{\bibfnamefont{T.~F.} \bibnamefont{Havel}},
  \bibinfo{author}{\bibfnamefont{M.~A.} \bibnamefont{Pravia}},
  \bibnamefont{and} \bibinfo{author}{\bibfnamefont{D.~G.} \bibnamefont{Cory}},
  \bibinfo{journal}{Phys. Rev. A} \textbf{\bibinfo{volume}{67}},
  \bibinfo{pages}{042322} (\bibinfo{year}{2003}).

\bibitem[{\citenamefont{Weinstein et~al.}(2004)\citenamefont{Weinstein, Havel,
  Emerson, Boulant, Saraceno, Lloyd, and Cory}}]{weinstein2004qpt}
\bibinfo{author}{\bibfnamefont{Y.}~\bibnamefont{Weinstein}},
  \bibinfo{author}{\bibfnamefont{T.}~\bibnamefont{Havel}},
  \bibinfo{author}{\bibfnamefont{J.}~\bibnamefont{Emerson}},
  \bibinfo{author}{\bibfnamefont{N.}~\bibnamefont{Boulant}},
  \bibinfo{author}{\bibfnamefont{M.}~\bibnamefont{Saraceno}},
  \bibinfo{author}{\bibfnamefont{S.}~\bibnamefont{Lloyd}}, \bibnamefont{and}
  \bibinfo{author}{\bibfnamefont{D.}~\bibnamefont{Cory}}, \bibinfo{journal}{J.
  Chem. Phys.} \textbf{\bibinfo{volume}{121}}, \bibinfo{pages}{6117}
  (\bibinfo{year}{2004}).

\bibitem[{\citenamefont{Mitchell et~al.}(2003)\citenamefont{Mitchell, Ellenor,
  Schneider, and Steinberg}}]{mitchell2003dpp}
\bibinfo{author}{\bibfnamefont{M.~W.} \bibnamefont{Mitchell}},
  \bibinfo{author}{\bibfnamefont{C.~W.} \bibnamefont{Ellenor}},
  \bibinfo{author}{\bibfnamefont{S.}~\bibnamefont{Schneider}},
  \bibnamefont{and} \bibinfo{author}{\bibfnamefont{A.~M.}
  \bibnamefont{Steinberg}}, \bibinfo{journal}{Phys. Rev. Lett.}
  \textbf{\bibinfo{volume}{91}}, \bibinfo{pages}{120402}
  (\bibinfo{year}{2003}).

\bibitem[{\citenamefont{Nambu and Nakamura}(2005)}]{nambu2005ein}
\bibinfo{author}{\bibfnamefont{Y.}~\bibnamefont{Nambu}} \bibnamefont{and}
  \bibinfo{author}{\bibfnamefont{K.}~\bibnamefont{Nakamura}},
  \bibinfo{journal}{Phys. Rev. Lett.} \textbf{\bibinfo{volume}{94}},
  \bibinfo{pages}{010404} (\bibinfo{year}{2005}).

\bibitem[{\citenamefont{Riebe et~al.}(2006)\citenamefont{Riebe, Kim, Schindler,
  Monz, Schmidt, K\"orber, H\"ansel, H\"affner, Roos, and Blatt}}]{Riebe:06}
\bibinfo{author}{\bibfnamefont{M.}~\bibnamefont{Riebe}},
  \bibinfo{author}{\bibfnamefont{K.}~\bibnamefont{Kim}},
  \bibinfo{author}{\bibfnamefont{P.}~\bibnamefont{Schindler}},
  \bibinfo{author}{\bibfnamefont{T.}~\bibnamefont{Monz}},
  \bibinfo{author}{\bibfnamefont{P.~O.} \bibnamefont{Schmidt}},
  \bibinfo{author}{\bibfnamefont{T.~K.} \bibnamefont{K\"orber}},
  \bibinfo{author}{\bibfnamefont{W.}~\bibnamefont{H\"ansel}},
  \bibinfo{author}{\bibfnamefont{H.}~\bibnamefont{H\"affner}},
  \bibinfo{author}{\bibfnamefont{C.~F.} \bibnamefont{Roos}}, \bibnamefont{and}
  \bibinfo{author}{\bibfnamefont{R.}~\bibnamefont{Blatt}},
  \bibinfo{journal}{Phys. Rev. Lett.} \textbf{\bibinfo{volume}{97}},
  \bibinfo{pages}{220407} (\bibinfo{year}{2006}).

\bibitem[{\citenamefont{Mohseni and Rezakhani}(2009)}]{mohseni2009emp}
\bibinfo{author}{\bibfnamefont{M.}~\bibnamefont{Mohseni}} \bibnamefont{and}
  \bibinfo{author}{\bibfnamefont{A.~T.} \bibnamefont{Rezakhani}},
  \bibinfo{journal}{Phys. Rev. A} \textbf{\bibinfo{volume}{80}},
  \bibinfo{pages}{010101} (\bibinfo{year}{2009}).

\bibitem[{\citenamefont{Kofman and Korotkov}(2009)}]{kofman2009}
\bibinfo{author}{\bibfnamefont{A.G.}~\bibnamefont{Kofman}} \bibnamefont{and}
  \bibinfo{author}{\bibfnamefont{A.N.} \bibnamefont{Korotkov}},
  \bibinfo{journal}{Phys. Rev. A} \textbf{\bibinfo{volume}{80}},
  \bibinfo{pages}{042103} (\bibinfo{year}{2009}).


\bibitem[{\citenamefont{Candes and Wakin}(2008)}]{candes2008phl}
\bibinfo{author}{\bibfnamefont{E.}~\bibnamefont{Candes}} \bibnamefont{and}
  \bibinfo{author}{\bibfnamefont{M.}~\bibnamefont{Wakin}},
  \bibinfo{journal}{IEEE Sig. Proc. Mag.} \textbf{\bibinfo{volume}{25}},
  \bibinfo{pages}{21} (\bibinfo{year}{2008}).

\bibitem[{\citenamefont{Candes and Tao}(2008)}]{candes2008rcs}
\bibinfo{author}{\bibfnamefont{E.~J.} \bibnamefont{Candes}} \bibnamefont{and}
  \bibinfo{author}{\bibfnamefont{T.}~\bibnamefont{Tao}}, \bibinfo{journal}{IEEE
  Inform. Theory News.} \textbf{\bibinfo{volume}{Dec. 14}}
  (\bibinfo{year}{2008}).

\bibitem[{\citenamefont{Kosut}(2008)}]{Kosut:08qpt}
\bibinfo{author}{\bibfnamefont{R.~L.} \bibnamefont{Kosut}},
  \bibinfo{journal}{arXiv:0812.4323v1[quant-ph]}  (\bibinfo{year}{2008}).

\bibitem[{\citenamefont{Katz et~al.}(2009)\citenamefont{Katz, Bromberg, and
  Silberberg}}]{katz2009cgi}
\bibinfo{author}{\bibfnamefont{O.}~\bibnamefont{Katz}},
  \bibinfo{author}{\bibfnamefont{Y.}~\bibnamefont{Bromberg}}, \bibnamefont{and}
  \bibinfo{author}{\bibfnamefont{Y.}~\bibnamefont{Silberberg}},
  \bibinfo{journal}{Applied Physics Letters} \textbf{\bibinfo{volume}{95}},
  \bibinfo{pages}{131110} (\bibinfo{year}{2009}).


\bibitem[{\citenamefont{Gross et~al.}(2010)\citenamefont{Gross, Liu, Flammia,
  Becker, and Eisert}}]{Gross:09}
\bibinfo{author}{\bibfnamefont{D.}~\bibnamefont{Gross}},
  \bibinfo{author}{\bibfnamefont{Y.-K.} \bibnamefont{Liu}},
  \bibinfo{author}{\bibfnamefont{S.~T.} \bibnamefont{Flammia}},
  \bibinfo{author}{\bibfnamefont{S.}~\bibnamefont{Becker}}, \bibnamefont{and}
  \bibinfo{author}{\bibfnamefont{J.}~\bibnamefont{Eisert}},
  \bibinfo{journal}{Phys. Rev. Lett.} \textbf{\bibinfo{volume}{105}},
  \bibinfo{pages}{150401} (\bibinfo{year}{2010}).

\bibitem[{\citenamefont{Candes}(2008)}]{Candes:08}
\bibinfo{author}{\bibfnamefont{E.~J.} \bibnamefont{Candes}},
  \bibinfo{journal}{Compte Rendus de l'Academie des Sciences, Paris, Serie I}
  \textbf{\bibinfo{volume}{346}}, \bibinfo{pages}{589} (\bibinfo{year}{2008}).

\bibitem[{\citenamefont{Needell and Tropp}(2008)}]{cosamp:08}
\bibinfo{author}{\bibfnamefont{D.}~\bibnamefont{Needell}} \bibnamefont{and}
  \bibinfo{author}{\bibfnamefont{J.~A.} \bibnamefont{Tropp}},
  \bibinfo{journal}{Appl. Comp. Harmonic Anal.} pp. \bibinfo{pages}{301--321}
  (\bibinfo{year}{2008}).

\bibitem[{\citenamefont{Baraniuk et~al.}(2010)\citenamefont{Baraniuk, Cevher,
  Duarte, and Hegde}}]{baraniuk2010mbc}
\bibinfo{author}{\bibfnamefont{R.~G.} \bibnamefont{Baraniuk}},
  \bibinfo{author}{\bibfnamefont{V.}~\bibnamefont{Cevher}},
  \bibinfo{author}{\bibfnamefont{M.~F.} \bibnamefont{Duarte}},
  \bibnamefont{and} \bibinfo{author}{\bibfnamefont{C.}~\bibnamefont{Hegde}},
  \bibinfo{journal}{IEEE Transactions on Information Theory}
  \textbf{\bibinfo{volume}{56}}, \bibinfo{pages}{1982} (\bibinfo{year}{2010}).

\bibitem[{\citenamefont{Yuen-Zhou et~al.}(2010)\citenamefont{Yuen-Zhou,
  Mohseni, and Aspuru-Guzik}}]{yuen2010qpt}
\bibinfo{author}{\bibfnamefont{J.}~\bibnamefont{Yuen-Zhou}},
  \bibinfo{author}{\bibfnamefont{M.}~\bibnamefont{Mohseni}}, \bibnamefont{and}
  \bibinfo{author}{\bibfnamefont{A.}~\bibnamefont{Aspuru-Guzik}},
  \bibinfo{journal}{Arxiv preprint arXiv:1006.4866}  (\bibinfo{year}{2010}).

\bibitem[{\citenamefont{Baraniuk et~al.}(2008)\citenamefont{Baraniuk,
  Davenport, DeVore, and Wakin}}]{Baraniuk:2008}
\bibinfo{author}{\bibfnamefont{R.}~\bibnamefont{Baraniuk}},
  \bibinfo{author}{\bibfnamefont{M.}~\bibnamefont{Davenport}},
  \bibinfo{author}{\bibfnamefont{R.}~\bibnamefont{DeVore}}, \bibnamefont{and}
  \bibinfo{author}{\bibfnamefont{M.}~\bibnamefont{Wakin}},
  \bibinfo{journal}{Constructive Approximation} \textbf{\bibinfo{volume}{28}},
  \bibinfo{pages}{253} (\bibinfo{year}{2008}).

\bibitem[{\citenamefont{Lorentz et~al.}(1996)\citenamefont{Lorentz, Golitschek,
  and Makovoz}}]{Lorentz}
\bibinfo{author}{\bibfnamefont{G.~G.} \bibnamefont{Lorentz}},
  \bibinfo{author}{\bibfnamefont{M.~v.} \bibnamefont{Golitschek}},
  \bibnamefont{and} \bibinfo{author}{\bibfnamefont{Y.}~\bibnamefont{Makovoz}},
  \bibinfo{journal}{Grundlehren der Mathematischen Wissenschaften}
  \textbf{\bibinfo{volume}{304}} (\bibinfo{year}{1996}).

\bibitem[{\citenamefont{Lofberg}(2004)}]{yalmip:04}
\bibinfo{author}{\bibfnamefont{J.}~\bibnamefont{Lofberg}}, in
  \emph{\bibinfo{booktitle}{Proceedings of the CACSD Conference}}
  (\bibinfo{address}{Taipei, Taiwan}, \bibinfo{year}{2004}).

\bibitem[{\citenamefont{Toh et~al.}(2004)\citenamefont{Toh, Tutuncu, and
  Todd}}]{Sdpt3}
\bibinfo{author}{\bibfnamefont{K.~C.} \bibnamefont{Toh}},
  \bibinfo{author}{\bibfnamefont{R.~H.} \bibnamefont{Tutuncu}},
  \bibnamefont{and} \bibinfo{author}{\bibfnamefont{M.~J.} \bibnamefont{Todd}}
  (\bibinfo{year}{2004}),
  \bibinfo{note}{http://www.math.nus.edu.sg/$\sim$mattohkc/sdpt3.html}.

\bibitem[{\citenamefont{Langford et~al.}(2005)\citenamefont{Langford, Weinhold,
  Prevedel, Resch, Gilchrist, O'Brien, Pryde, and White}}]{langford2005dse}
\bibinfo{author}{\bibfnamefont{N.~K.} \bibnamefont{Langford}},
  \bibinfo{author}{\bibfnamefont{T.~J.} \bibnamefont{Weinhold}},
  \bibinfo{author}{\bibfnamefont{R.}~\bibnamefont{Prevedel}},
  \bibinfo{author}{\bibfnamefont{K.~J.} \bibnamefont{Resch}},
  \bibinfo{author}{\bibfnamefont{A.}~\bibnamefont{Gilchrist}},
  \bibinfo{author}{\bibfnamefont{J.~L.} \bibnamefont{O'Brien}},
  \bibinfo{author}{\bibfnamefont{G.~J.} \bibnamefont{Pryde}}, \bibnamefont{and}
  \bibinfo{author}{\bibfnamefont{A.~G.} \bibnamefont{White}},
  \bibinfo{journal}{Phys. Rev. Lett.} \textbf{\bibinfo{volume}{95}},
  \bibinfo{pages}{21} (\bibinfo{year}{2005}).

\bibitem[{\citenamefont{Okamoto et~al.}(2005)\citenamefont{Okamoto, Hofmann,
  Takeuchi, and Sasaki}}]{okamoto2005doq}
\bibinfo{author}{\bibfnamefont{R.}~\bibnamefont{Okamoto}},
  \bibinfo{author}{\bibfnamefont{H.~F.} \bibnamefont{Hofmann}},
  \bibinfo{author}{\bibfnamefont{S.}~\bibnamefont{Takeuchi}}, \bibnamefont{and}
  \bibinfo{author}{\bibfnamefont{K.}~\bibnamefont{Sasaki}},
  \bibinfo{journal}{Phys. Rev. Lett.} \textbf{\bibinfo{volume}{95}},
  \bibinfo{pages}{210506} (\bibinfo{year}{2005}).

\bibitem[{\citenamefont{Kiesel et~al.}(2005)\citenamefont{Kiesel, Schmid,
  Weber, Ursin, and Weinfurter}}]{kiesel2005loc}
\bibinfo{author}{\bibfnamefont{N.}~\bibnamefont{Kiesel}},
  \bibinfo{author}{\bibfnamefont{C.}~\bibnamefont{Schmid}},
  \bibinfo{author}{\bibfnamefont{U.}~\bibnamefont{Weber}},
  \bibinfo{author}{\bibfnamefont{R.}~\bibnamefont{Ursin}}, \bibnamefont{and}
  \bibinfo{author}{\bibfnamefont{H.}~\bibnamefont{Weinfurter}},
  \bibinfo{journal}{Phys. Rev. Lett.} \textbf{\bibinfo{volume}{95}},
  \bibinfo{pages}{210505} (\bibinfo{year}{2005}).

\bibitem[{\citenamefont{Weinhold et~al.}(2008)\citenamefont{Weinhold,
  Gilchrist, Resch, Doherty, O'Brien, Pryde, and White}}]{weinhold2008upq}
\bibinfo{author}{\bibfnamefont{T.}~\bibnamefont{Weinhold}},
  \bibinfo{author}{\bibfnamefont{A.}~\bibnamefont{Gilchrist}},
  \bibinfo{author}{\bibfnamefont{K.}~\bibnamefont{Resch}},
  \bibinfo{author}{\bibfnamefont{A.}~\bibnamefont{Doherty}},
  \bibinfo{author}{\bibfnamefont{J.}~\bibnamefont{O'Brien}},
  \bibinfo{author}{\bibfnamefont{G.}~\bibnamefont{Pryde}}, \bibnamefont{and}
  \bibinfo{author}{\bibfnamefont{A.}~\bibnamefont{White}},
  \bibinfo{journal}{Preprint at arXiv:0808.0794}  (\bibinfo{year}{2008}).

\bibitem[{\citenamefont{Barbieri et~al.}(2009)\citenamefont{Barbieri, Weinhold,
  Lanyon, Gilchrist, Resch, Almeida, and White}}]{barbieri2009pdo}
\bibinfo{author}{\bibfnamefont{M.}~\bibnamefont{Barbieri}},
  \bibinfo{author}{\bibfnamefont{T.}~\bibnamefont{Weinhold}},
  \bibinfo{author}{\bibfnamefont{B.}~\bibnamefont{Lanyon}},
  \bibinfo{author}{\bibfnamefont{A.}~\bibnamefont{Gilchrist}},
  \bibinfo{author}{\bibfnamefont{K.}~\bibnamefont{Resch}},
  \bibinfo{author}{\bibfnamefont{M.}~\bibnamefont{Almeida}}, \bibnamefont{and}
  \bibinfo{author}{\bibfnamefont{A.}~\bibnamefont{White}},
  \bibinfo{journal}{Journal of Modern Optics} \textbf{\bibinfo{volume}{56}},
  \bibinfo{pages}{209} (\bibinfo{year}{2009}).

\end{thebibliography}

\end{document}